\documentclass[11pt]{article}
\usepackage[margin=0.95in]{geometry}
\usepackage{amsmath, amsthm, amssymb}
\usepackage{times,esint,stackrel}
\usepackage{empheq}
\usepackage{enumitem}
\usepackage{color}  
\usepackage[colorlinks=true,, linkcolor=blue, citecolor=blue, urlcolor=blue]{hyperref}
\usepackage{mathtools}
\usepackage{etoolbox}
\usepackage{mathrsfs}
\usepackage{framed}
\usepackage[titles]{tocloft}

\usepackage{graphicx}
\usepackage[page]{appendix}
\allowdisplaybreaks
\makeatletter
\patchcmd\@thm
{\let\thm@indent\indent}{\let\thm@indent\noindent}
{}{}
\makeatother

\let\OLDthebibliography\thebibliography
\renewcommand\thebibliography[1]{
	\OLDthebibliography{#1}
	\setlength{\parskip}{0pt}
	\setlength{\itemsep}{0pt plus 0.1ex}
}

\numberwithin{equation}{section}

\newcommand{\be}{\begin{equation}}
\newcommand{\ee}{\end{equation}}
\newcommand{\bea}{\begin{eqnarray}}
\newcommand{\eea}{\end{eqnarray}}

\newtheorem{thm}{Theorem}

\theoremstyle{definition}
\newtheorem{remark}{Remark}
\newtheorem{definition}{Definition}

\newcommand{\ve}{{\varepsilon}}

\newcommand{\rmd}{{\rm d}}

\newcommand{\bE}{{\mathbb{E}}}

\newcommand{\bq}{\begin{equation}}
\newcommand{\eq}{\end{equation}}
\newcommand{\bqa}{\begin{eqnarray*}}
\newcommand{\eqa}{\end{eqnarray*}}

\newcommand{\scp}[2]{{\big\langle {#1}\, , \, {#2}\big\rangle}}
\newcommand{\Scp}[2]{{\Big\langle {#1}\, , \, {#2}\Big\rangle}}
\newcommand{\SCP}[2]{{\left\langle {#1}\, , \, {#2}\right\rangle}}

\newcommand{\mb}[1]{\mbox{\boldmath{$#1$}}}

\newcommand{\E}[1]{\mathbb{E}\left[{#1}\right]}

\mathtoolsset{showonlyrefs,showmanualtags}

\title{\vspace{-7mm}{ \bf \Large Lagrangian averaged stochastic advection by Lie transport for fluids}}
\author{
{ \bf Theodore D. Drivas} \thanks{\footnotesize Department of Mathematics, Princeton University, Princeton, NJ 08544, \href{mailto:tdrivas@math.princeton.edu}{tdrivas@math.princeton.edu}}
\and
{ \bf Darryl D. Holm} \thanks{\footnotesize Department of Mathematics, Imperial College, London SW7 2AZ, UK, \href{mailto:d.holm@imperial.ac.uk }{d.holm@imperial.ac.uk}}
\and
{ \bf James-Michael Leahy} \thanks{\footnotesize Department of Mathematics, Imperial College, London SW7 2AZ, UK, \href{mailto:j.leahy@imperial.ac.uk}{j.leahy@imperial.ac.uk}}
}
\date{} %
\begin{document}
\maketitle
\vspace{-5mm}
\begin{abstract} 
We formulate a class of stochastic partial differential equations based on Kelvin's circulation theorem for ideal fluids. In these models, the velocity field is randomly transported by white-noise vector fields, as well as by its own average over realizations of this noise. We call these systems  the \emph{Lagrangian averaged stochastic advection by Lie transport} (LA SALT) equations. These equations are nonlinear and non-local, in both physical and probability space.  Before taking this average, the equations recover the Stochastic Advection by Lie Transport (SALT) fluid equations introduced by Holm \cite{Holm:2015aa}.  Remarkably, the introduction of the non-locality in probability space in the form of momentum transported by its own mean velocity gives rise to a closed equation for the expectation field which comprises Navier--Stokes equations with Lie--Laplacian `dissipation'.   As such, this form of non-locality provides a regularization mechanism.  The formalism we develop is closely connected to the stochastic Weber velocity framework of Constantin and Iyer \cite{Constantin:2008aa}  in the case when the noise correlates are taken to be the constant basis vectors in $\mathbb{R}^3$ and, thus, the Lie--Laplacian reduces to the usual Laplacian. 
We extend this class of equations to allow for advected quantities to be present and affect the flow through exchange of kinetic and potential energies. 
The statistics of the solutions for the LA SALT fluid equations are found to be changing  dynamically due to an array of intricate correlations among the physical variables. 
The statistical properties of the LA SALT physical variables propagate as local evolutionary equations which when spatially integrated become dynamical equations for the variances of the fluctuations. 
Essentially, the LA SALT theory is a non-equilibrium stochastic linear response theory for fluctuations in SALT fluids with advected quantities.
\end{abstract}

\renewcommand{\baselinestretch}{0.8}\normalsize
\tableofcontents
\renewcommand{\baselinestretch}{1.0}\normalsize

\section{Introduction -- SALT and LA SALT}
In honour of the 150$^{\rm th}$ anniversary of the publication of Kelvin's circulation theorem for Euler's fluid equations of deterministic incompressible flow, Drivas and Holm \cite{Drivas:2018aa} derived two classes of \emph{stochastic} Euler fluid equations, distinguished by whether they either preserve circulation as in Holm \cite{Holm:2015aa}, or they preserve kinetic energy, as in \cite{mikulevicius2004stochastic} and \cite{flandoli2011interaction}. The paper  \cite{Drivas:2018aa} also introduced the mean field equation ``in the sense of McKean'' \cite{McKean:1966aa} for stochastic Euler fluid dynamics which preserves a statistical circulation theorem.  The resulting mean field equation in paper \cite{Drivas:2018aa} was related to the results of Constantin and Iyer \cite{Constantin:2008aa}, in which smooth solutions of Navier--Stokes are shown to be characterized by a special case of this statistical Kelvin's theorem. The present paper extends the mean field approach of \cite{Drivas:2018aa} to include stochastic fluid equations with potential energy whose deterministic version falls in the geometric class of fluid equations studied in Holm, Marsden, and Ratiu \cite{Holm:1998aa}. These stochastic partial differential equation systems (SPDEs) comprise the equations of  \emph{Lagrangian Averaged Stochastic Advection by Lie Transport} (LA SALT) corresponding to the deterministic fluid equations treated in \cite{Holm:1998aa}.  Besides the mean field, solutions of the LA SALT equations also contain the time-dependent statistics of the stochastic fluctuations around the mean field solutions. Thus, the LA SALT formulation may be regarded as a stochastic version of linear response theory.

\subsection{Purpose}
This paper develops a new class of fluid equations based on Stochastic Advection by Lie Transport (SALT) \cite{Holm:2015aa} by applying a type of Lagrangian Average (LA) which is the counterpart in probability space of the time average at fixed Lagrangian coordinate taken in the LANS-alpha turbulence model \cite{Foias:2001aa, Foias:2002aa}. As was proven previously for SALT, the new set of LA SALT fluid equations with advected quantities preserves the fundamental properties of ideal fluid dynamics. These properties include Kelvin's circulation theorem and Lagrangian invariants such as enstrophy, helicity and potential vorticity  \cite{Holm:2015aa}. These properties derive from the preserved Euler--Poincar\'e and Lie--Poisson geometric structures of the deterministic theory, whose history and basic mathematical elements are reviewed briefly in appendices \ref{Arnold-sec} and \ref{sec:GeomSet} for the convenience of the reader. In the Kelvin circulation integral for the new theory (LA SALT), the Lagrangian Average \emph{in probability space} manifests itself as advection of the Kelvin circulation loop by the \emph{expected} fluid transport velocity. This is the primary modification made by applying LA to the SALT theory, \cite{Holm:2015aa, Cotter:2017aa, Cotter:2018aa, Leon:2019aa, Cotter:2019aa, Cotter:2019ab}. This modification of the SALT theory allows the dynamics of the statistical properties of the solutions for the LA SALT fluid equations to be investigated directly.  

One interesting consequence is that, due to the non-local nature of the equations in probability space, the expected dynamics contains terms which regularize the (expected) solution. Thus, the introduction of the LA modification of Kelvin's circulation integral bestows on the LA SALT version of the 3D Euler fluid   similar solution properties to those of the incompressible Navier-Stokes equations. 
Moreover, in certain cases, the additional terms are strictly dissipative, and the LA SALT fluid equations may be regarded as non-conservative system of PDEs for the expected motion embedded into a larger conservative system which includes the fluctuation dynamics.  From this viewpoint, the interaction dynamics of the two components of the full LA SALT system dissipates the Lagrangian invariant functions of the mean quantities by converting them into fluctuations, while preserving the  total  invariants. We will explore these properties both at the level of a general semidirect-product Hamiltonian system and more concretely for special cases of such systems where more details can be worked out.

\vspace{1mm} \noindent {\bf Plan of the paper.} 
We write the expected-quantity equations for LA SALT fluid dynamics with advected quantities in the semidirect-product Hamiltonian matrix form of equation \eqref{eq:LASALT-EP} in Section \ref{sec:LiePoisHam}. We then point out that the fluctuations are slaved to the expected equations in a certain sense. This slaving relation enables us to calculate the evolution equations for the local and spatially integrated variances of the fluctuations. 

Section \ref{sec:LASALT-Eu} begins by discussing the solution properties of vorticity dynamics and helicity conservation in the example of the LA SALT Euler equations for ideal fluids. 
In section \ref{Prob-sec}, we then discuss the mathematical well-posedness of the LA SALT Euler fluid equations.  Theorem 1 establishes local existence for the LA SALT Euler fluid equations in Sobolev spaces in $d=2,3$, as well as global existence provided we work in two-spatial dimensions or the magnitudes of the  noise correlates $\{\xi^{(k)}\}_k$ are sufficiently `large' relative to the data and forces in three-dimensions. To establish these results, we take advantage of the fact the dynamics of the expectation field decouples from the fluctuations and solves a closed Navier-Stokes type equation with Lie-Laplacian  `dissipation' (LL NS).  Theorem 2 establishes the results above for LL NS. 

Section \ref{sec:examples} illustrates the breadth of the LA SALT theory by discussing several examples. It begins by treating three simple examples which do not have advected quantities. These examples comprise the rigid body in 3D, the Burgers equation, and the Camassa-Holm equation. These first three examples reveal the additional information made available in the LA SALT theory for investigating both the linear transport equations for the fluctuation dynamics and the evolution equations for the expected physical variables, such as variances. In conclusion, we demonstrate how a more advanced example fits into the LA SALT theory. For this purpose, we formulate the 3D LA SALT incompressible stratified magnetohydrodynamics (MHD) equations and point out that MHD at this level also contains 3D Euler--Boussinesq, whose 2D case has been treated in the companion paper \cite{abht}.

\vspace{1mm}\noindent{\bf Main content of the paper:}
\begin{itemize}
\item
Stochastic Euler fluid equations which are ``nonlinear in the sense of McKean'' \cite{McKean:1966aa, Constantin:2008aa, Drivas:2018aa, Hochgerner:2018aa} are generalised to include advected quantities. 
\item
In certain cases, the dynamical  equations for the expected values of physical variables decouple as a subsystem from the fluctuation dynamics. In the absence of advected quantities, a particular case yields an extension of the Navier--Stokes partial differential equations (PDE).  
\item
The SPDEs for the dynamics of the fluctuations of the momentum and advected quantities are shown to be transported by the PDE solutions for the expected values.
\item
The statistical properties of the fluctuations are found to evolve dynamically, driven by an intricate array of correlations. Specifically, the statistical properties of the LA SALT physical variables propagate as local equations and yield dynamical variances when spatially integrated. In certain cases for which the dynamics of the fluctuations occurs by linear transport, the equations for their statistical properties form closed evolutionary PDE systems.
\item
Analytical conditions are found for which the LA SALT Euler fluid equations (a specific example of the LA SALT equations) are well-posed. 
\end{itemize}

For the reader's convenience, the historical context and the basic mathematical elements of our present geometrical approach to fluid dynamics are briefly surveyed in Appendix \ref{Arnold-sec} and \ref{sec:GeomSet}, respectively.

\vspace{1mm}\noindent{\bf Flow interpretation of Stochastic Advection by Lie Transport (SALT).}
In a paper in 2015, Holm \cite{Holm:2015aa} extended the Clebsch approach of \cite{Holm:1983aa}  for deriving Euler--Poincar\'e equations for fluids with advected quantities to the case that the fluid variables undergo flows on Diff or SDiff generated via Stochastic Advection by Lie Transport (SALT). The papers \cite{Holm:2015aa,Gay-Balmaz:2018aa} derived the SALT class of stochastic continuum equations from Hamilton's principle with a stochastic advection constraint on Lagrangian fluid trajectories. Later, the stochastic constraint in \cite{Holm:2015aa} was derived in \cite{Cotter:2017aa} from a corresponding two-scale decomposition of the deterministic fluid flow map using a deterministic homogenisation procedure.  In \cite{Cotter:2017aa}, the  fluid flow map 
\begin{align}
x_t:={g}_t x_0\in M
\quad\hbox{with}\quad
{g}_0 x_0=x_0
\,
\end{align}
is decomposed into a composition of two time-dependent diffeomorphisms
\begin{align}\label{dx-form}
g_{t}= \widetilde{g}_{t/\epsilon}\circ  \bar{g}^{\epsilon}_t = (\operatorname{Id} + \gamma_{t/\epsilon})\circ \bar{g}^{\epsilon}_t\,.
\end{align}
The above formula factorises the flow map $g_{t}$ into the composition of two maps with different time scales $t$ and $t/\epsilon$ which are well-separated when $\epsilon\ll1$.  Let $u_t = \dot{g}_tg_t^{-1}$ and define  $\bar{x}^{\epsilon}_t(x_0)=\bar{g}^{\epsilon}_t x_0$ to be the  fluid trajectory associated with the slow time.  It follows from  the decomposition \eqref{dx-form} and the chain rule that
$$
\left(\operatorname{Id} + \nabla_{\bar{x}_t^{\epsilon}}\gamma(\bar{x}^{\epsilon}_t(x_0),t/\epsilon)\right) \dot{\bar{x}}_t^{\epsilon}(x_0)=u_t ( \bar{g}_t^{\epsilon}  x_0 + \gamma_{t/\epsilon} \bar{g}_t^{\epsilon} x_0 )- \frac{1}{\epsilon}\, \frac{\partial}{\partial(t/\epsilon)}\gamma (\bar{x}_t^{\epsilon}(x_0),t/\epsilon),
$$
where $\dot{\bar{x}}_t^{\epsilon}(x_0)$ is the the time-derivative of $\bar{x}^{\epsilon}_t(x_0)$ at fixed $x_0$. 

Deterministic multi-time homogenisation theory was used in  \cite{Cotter:2017aa} to show that $\bar{x}^{\epsilon}(x_0)=\bar{g}^{\epsilon} x_0$ tends to $\bar{x}(x_0)=\bar{g}x_0$ in the limit as $\epsilon\rightarrow 0$,  where $\bar{x}_t(x_0)$ is the solution of the following It\^o stochastic differential equation
\begin{align}\label{SDEbardx}
\rmd \bar{x}_t(x_0) =\bar{u}_t (\bar{x}_t(x_0))\rmd t + \sum_k \sigma(\bar{x}_t(x_0)) \rmd W_t\,,
\end{align}
provided two additional conditions are satisfied. First, one requires that $\partial_t \gamma$ has a positive Lyapunov exponent, so there exists a compact attractor and an ergodic invariant probability measure $\mu$  associated with $\partial_t \gamma$ arising from sensitivity to initial conditions. Second, one requires a centering condition $\int (\operatorname{Id}+\nabla \gamma)^{-1}u d\mu=0$ to hold, so the fluctuating flow velocity has zero mean when pulled back to the mean flow.

\begin{remark}[The SALT theory]
The SALT theory begins with the stochastic flow map $\phi_t$ generated by the following Stratonovich stochastic vector field equivalent to \eqref{SDEbardx}\footnote{We use the notation 
$$
\rmd x_t(x) =b_t (x)\rmd t + \sum_k \xi^{(k)}(x)\circ \rmd W_t^{(k)}, \;\; x\in M,
$$
to denote the \textit{stochastic vector-field} associated with the  stochastic flow $\phi=\{\phi_{s,t}\}_{0\leq s\leq t}$:
$$
\rmd\phi_{s,t}(x)=b_t (\phi_{s,t}(x))\rmd t +  \sum_k \xi^{(k)}(\phi_{s,t}(x))\circ \rmd W_t, \quad \phi_{s,s}(x)=x\in M.
$$
That is to say, $\rmd x_t=\rmd \phi_{0,t} \circ \phi^{-1}_{0,t}$ is the stochastic analogue of the usual Eulerian vector field.
For a given tensor-field $\tau$ on $M$, we  use the notation $\pounds_{\rmd x_t} \tau$ to denote the Lie derivative of $\tau$ along the stochastic vector field $\rmd X_t$. This may be defined simply as 
$\pounds_{ \rmd x_t} \tau= \pounds_{b_t} \tau  \rmd t + \pounds_{\xi^{(k)}} \tau  \circ \rmd W_t^{(k)},$
or, perhaps, more satisfactorily, as $\pounds_{ \rmd  x_t} \tau=\rmd_{\epsilon} \phi_{t,t+\epsilon}^*\tau |_{\epsilon=0}$, which follows from the It\^o-Kunita-Wentzell formula (see, e.g.,  \cite{Leon:2019aa}). This notation generalizes the definition of the deterministic  Lie-derivative to the stochastic case.
}
\begin{align}\label{dx-form1}
\rmd x_t(x)= u^L_t (x) \rmd t+ \sum_{k=1}^{\infty} \xi^{(k)}(x)\circ \rmd W_t^{(k)}, \;\;x\in M
\,,
\end{align}
where the drift $u^L_t$ may be random and depend on the noise in \eqref{dx-form1}.  

Although the premise of the flow in equation \eqref{dx-form} involves a separation of time scales, as we have discussed, the derivation of the stochastic vector field \eqref{SDEbardx} or its equivalent  \eqref{dx-form1} by using homogenisation in time introduces an assumption of ergodicity based on a positive Lyapunov exponent at faster time scales \cite{Cotter:2017aa}. Stochastic computational simulations of SALT fluid flows using the stochastic vector field \eqref{dx-form1} can then possess broad-band temporal spectra whose ensemble statistics is non-Gaussian, rather than showing a separation of time sales.  See \cite{Cotter:2019aa, Cotter:2019ab} for details.

In paper \cite{Holm:2015aa}, the drift velocity of \eqref{dx-form1} was determined using the Clebsch variational approach in Hamilton's principle for fluid dynamics. This approach ensured that the Lagrangian invariants of the deterministic flow would still hold in each realization of the process \eqref{dx-form1}.  This is the SALT approach, in which the vector field \eqref{dx-form1} appears in the stochastic perturbation of the Euler--Poincar\'{e} equations with Lie transport noise. In applications of the SALT approach, the stationary vector fields $\xi^{(k)}(x)$ have been determined from a data analysis procedure involving numerical simulations of the stochastic equations of motion. Such a procedure has been successfully developed in \cite{Cotter:2018aa, Cotter:2019aa, Cotter:2019ab}. Here, we will assume that these stationary vector fields have already been obtained from the appropriate data analysis for each given application.
\end{remark}

\noindent{\bf Aim of the present paper.}
The present paper aims to develop the following Lagrangian Average (LA) SALT class of stochastic continuum equations. Namely, we will consider stochastic flows $\{\Phi_t\}$ generated by stochastic vector fields of the form,
\begin{align}\label{dX-form}
\rmd X_t(x)= \E{u^L_t} (x)\, \rmd t+ \sum_{k=1}^{\infty} \xi^{(k)}(x)\circ \rmd W_t^{(k)}, \ \ x\in M \,,
\end{align}
when $\E{u^L_t}$ denotes the expectation of the velocity vector field $u^L_t$ with respect to the stochastic flow map $\phi_t$ generated by the SALT vector field in \eqref{dx-form1}.
It may be possible to interpret this choice as a further averaging at the homogenisation level, but we will leave this supposition for future work. For now, we simply adopt the LA SALT formulation implied by the stochastic vector field \eqref{dX-form} and explore its dynamical consequences. 

The LA SALT formulation will be closed here by enforcing preservation of the same Lagrangian invariants as in  \cite{Holm:2015aa}, but now along the flow \eqref{dX-form}. The resulting equations are nonlinear in the sense of McKean, i.e. there is a non-locality in probability space (an expectation) taken in the nonlinear term. As we will see, this modification introduces new regularizing terms into the equations for the expected dynamical variables.  

\subsection{Kelvin theorem interpretation of SALT and LA SALT}

The modelling approach of Stochastic Advection by Lie Transport (SALT) \cite{Cotter:2017aa, Cotter:2018aa, Leon:2019aa, Cotter:2019aa, Cotter:2019ab} may be defined by enforcing a modification of the Kelvin theorem for deterministic fluids which replaces the transport velocity of the circulation  loop $u^L_t$ in the deterministic Kelvin theorem by a Stratonovich stochastic vector field $\rmd x_t$ whose drift velocity is the same as the circulation loop velocity $u^L_t$ \cite{Holm:2015aa, Drivas:2018aa}. That is, for a given smooth manifold $M$:
\begin{align}\label{SALT-Kel}
\oint_{C(u^L_t)} u_t\cdot d x
\quad\to\quad 
\oint_{C(\rmd x_t)}
u_t\cdot d x,
\end{align}
where $\rmd x_t$ is the SALT stochastic vector field in \eqref{dx-form1}. In the transition to the SALT Kelvin circulation integral in \eqref{SALT-Kel}, the notation $C(u_t^L)$ (resp.\ $C(\rmd x_t)$) is used to denote that  the closed material loop  at time $t$ that is moving along the flow associated with $u^L$ (resp.\ $\rmd x_t$) after having started at time $t=0$.  In particular, we have $C(\rmd x_t)=\phi_t(C)$, which is a family of  loops moving with the SALT flow $\phi_t$  at time $t$. 

For example, in the Euler fluid case the proof of the  modified stochastic Kelvin theorem for SALT  \cite{Leon:2019aa} is given by
\begin{align}\label{KelThm-Eul-SALT}
\begin{split}
{\rm d} \oint_{\phi_t(C)} u_t  \cdot d x
&=  \oint_{C} \rmd \big(\phi_t^* (u_0\cdot d x_0)\big)
\\&=\oint_{C} \phi_t^*
\big[ {\rm d} u_0\cdot d x_0 +  \pounds_{\rmd x_0}( u_0 \cdot d x_0) \big]
\\&=\oint_{\phi_t(C)}
\big[ {\rm d} u_t \cdot d x+  \pounds_{\rmd x_t}( u_t \cdot d x) \big]\\
&=-\oint_{\phi_t(C)} \nabla \rmd p_t\cdot dx 
= 0 \,,
\end{split}
\end{align}
where $\phi_t^*$ is the pull-back of the flow $\phi_t$,  $\pounds_{\rmd x_t}$ is the Lie derivative along the stochastic vector field $\rmd x_t$ which generates the flow $\phi_t$, and $\rmd  p_t$ is the pressure semimartingale for the SALT Euler fluid equation \cite{Crisan:2019aa}
\[
{\rm d} u_t \cdot d x+  \pounds_{\rmd x_t}( u_t \cdot d x)  
= - \nabla \rmd p_t \cdot d x
\quad\hbox{where}\quad
{\rm div}u_t  = 0
.
\]  
We refer the reader to \cite{Leon:2019aa} for a rigorous discussion of the calculation in \eqref{KelThm-Eul-SALT} and its association with Newton's Force Law for stochastic fluids. 
For more discussion of the emergence of Lie derivatives in the proof of Kelvin's circulation theorem as sketched in \eqref{KelThm-Eul} for fluid flow, one may also refer to  \cite{Holm:1998aa} in the deterministic case.

The same stochastic transport velocity $\rmd x_t$ in \eqref{dx-form1} which transports the circulation loop also advects the Lagrangian parcels in the SALT theory. The  Lagrangian parcels may carry advected quantities $a$, such as heat, mass and magnetic field lines, by Lie transport along with the flow, as
\begin{align}\label{advec-qty}
{\rm d} a + \pounds_{{\rm d} x_t }a = 0\,,
\end{align}
where  $\pounds_{{\rm d} x_t }a$ is the Lie derivative of a tensor field $a$ with respect to the vector field ${\rm d} x_t$ in equation \eqref{dx-form1}. That is, an advected tensor-field $a$ satisfies
\begin{align}\label{advec-def}
{\rm d} \big(\phi_t^*a(t,x)\big) =\phi_t^* \Big({\rmd } a(t,x) 
+ \mathcal{L}_{{\rm d} x_t} a(t,x)\Big) = 0
\,,\quad\hbox{a.s.}
\end{align}
where $\phi_t^*a=a_0$ is the pull-back of $a$ by the map $\phi_t$.
Formula \eqref{advec-def} defines advection as ``invariance under the SALT flow''. 
We refer the reader to \cite{Leon:2019aa} for more details about stochastic advection. 

In this paper, we modify the SALT approach to stochastic fluid dynamics by replacing the SALT map $\phi_t$ by the LA SALT map $\Phi_t$. Correspondingly, we replace the Eulerian vector field $\rmd x_t(x)$ in \eqref{dx-form1} by the vector field $\rmd X_t(x)$ in \eqref{dX-form}. Because the drift velocity will be replaced by the expected velocity, this replacement is reminiscent of the McKean--Vlasov mean field approach for finite dimensional stochastic flow, which replaces the velocities for an interacting particle system by their empirical mean \cite{Jabin:2017aa}.

The LA SALT approach modifies the SALT Kelvin circulation in \eqref{SALT-Kel} by replacing the drift velocity in the stochastic transport loop velocity in \eqref{dx-form1} by its expectation, plus the same noise as in SALT. Namely, \cite{Holm:2015aa, Drivas:2018aa}
\begin{align}\label{KelThm-form}
\oint_{C({\rm d} x_t)} u_t  \cdot dx
\quad\to \quad
\oint_{C({\rm d} X_t )} u_t \cdot dx\,,
\end{align}
where the stochastic vector field $\rmd X_t$ is given in \eqref{dX-form}.
Since the expectation in \eqref{dX-form} refers to the transport velocity $u^L_t$ of the Lagrangian loop in Kelvin's theorem, we regard this process as a probabilistic type of Lagrangian Average (LA) which is the counterpart of the time average at fixed Lagrangian coordinate taken in the LANS-alpha turbulence model \cite{Foias:2001aa, Foias:2002aa}. 

For example, in the Euler fluid case the modified Kelvin theorem reads (cf. equation \eqref{KelThm-Eul-SALT}),
\begin{align}\label{KelThm-Eul}
{\rm d} \oint_{C\big(\rmd X_t\big)} u_t  \cdot d x
=  
\oint_{C\big(\rmd X_t\big)}
\big[ {\rm d} u_t \cdot d x+  \pounds_{\rmd X_t}( u_t \cdot d x) \big]
= 0 \,,
\end{align}
where $ \pounds_{ {\rm d}X_t}( u_t \cdot  d x) $ denotes the Lie derivative of the 1-form 
$ u_t \cdot d x$ with respect to the vector field $ {\rm d}X_t$ given in equation \eqref{dX-form}.

\vspace{1mm}\noindent \textbf{LA SALT Euler equations.}
Evaluating equation \eqref{KelThm-Eul} implies the following stochastic Euler fluid motion equation in Stratonovich form
\begin{align}\label{LASALT-eqn}
{\rm d} u_t \cdot d x+   \mathbf{P}^\bE \pounds_{\rmd X_t}( u_t \cdot d x) = 0
\,,
\end{align}
where the projector $\mathbf{P}^\bE$ keeps the expectation of the transport velocity ($\E{u^L_t}$) divergence-free.   Requiring the expected transport velocity to be divergence-free and allowing the fluctuation field to be compressible will turn out to endow the LA SALT equations with the same Lie--Poisson Hamiltonian structure as that of both the SALT equations and their deterministic counterparts.  Moreover,  applying only the projector $\mathbf{P}^\bE$ in equation \eqref{LASALT-eqn} is the minimal assumption needed for the flow generated by $dX_t$ to be volume preserving, since our $\xi$'s are taken to be divergence-free. 

 This point will be detailed in Section \ref{sec:LiePoisHam}. 
The equivalent It\^o form of the Stratonovich equation \eqref{LASALT-eqn} is
\begin{align}\label{LASALT-eqn-Ito}
{\rm d} u_t\cdot d x+  \mathbf{P}^\bE\pounds_{\rmd\widehat{X}_t}( u_t\cdot d x) 
- \frac{1}{2}  \sum_k \mathbf{P}^\bE \pounds_{\xi^{(k)}}\big( \pounds_{\xi^{(k)}} ( u_t\cdot d x)\big) = 0
\,,
\end{align}
with It\^o vector field $ {\rm d}\widehat{X}_t$ given by 
\begin{align}\label{dXhat-form}
{\rm d}\widehat{X}_t(x) = \E{u^L_t}(x)\rmd t+ \sum_k \xi^{(k)}(x) \rmd W_t^{(k)}
\,.
\end{align}
Equations are said to be  ``nonlinear in the sense of McKean'' \cite{McKean:1966aa, Constantin:2008aa, Hochgerner:2018aa} when  the drift term involves the expected value of the flow it drives, as occurs in the pair of stochastic differential equations for the stochastic Euler fluid in \eqref{LASALT-eqn-Ito} and \eqref{dXhat-form}. This perspective is also adopted by \cite{Hochgerner:2018aa} in the special case of incompressible fluid and is used there as a route towards obtained a representation theorem for solutions of the deterministic incompressible Navier-Stokes equations. 

The expectation of the It\^o form of the Euler fluid motion equation \eqref{LASALT-eqn-Ito} yields a motion equation with additional terms. This is the Navier--Stokes equation with Lie--Laplacian `dissipation' \cite{Drivas:2018aa},
\begin{align}\label{LLNS-eqn}
\partial_t\E{u_t}\cdot d x+ \mathbf{P}   \pounds_{\E{u^L_t}}( \E{u_t}\cdot dx) 
- \frac{1}{2}  \sum_k \mathbf{P} \pounds_{\xi^{(k)}}\big( \pounds_{\xi^{(k)}} \big(\E{u_t} \cdot d x\big)   
=0
\,,
\end{align}
where $\mathbf{P}$ denotes the standard Leray projector of the vector coefficients onto their divergence-free part.  Note that, in general, the Lie--Laplacian operator is not dissipative. However, it is a uniformly elliptic operator which has the effect of regularizing the solution, provided the noise correlates satisfy certain minor conditions. The LLNS equation \eqref{LLNS-eqn} reduces to the Navier--Stokes equation for the special choice of the noise correlates  $ \xi^{(k)}=\{(1,0,0)^T,(0,1,0)^T,(0,0,1)^T\}$, for $k=1,2,3$. 

The LA SALT equations \eqref{LASALT-eqn-Ito} and \eqref{LLNS-eqn} first arose, along with additional noisy and viscous terms, in Lemma 3 of \cite{Drivas:2018aa} where they were shown to govern the dynamics of the so-called `stochastic Weber velocity'. Constantin and Iyer  \cite{Constantin:2008aa} also used stochastic fluid motion equations of the type \eqref{LASALT-eqn-Ito}--\eqref{dXhat-form} with \emph{constant} $\xi^{(k)}$, where $k=1,2,3,$ as a tool to represent solutions to the incompressible Navier-Stokes equations as an average over a stochastic process. In particular, in the special case of constant $\xi$'s, the following statistical Kelvin theorem was derived in  \cite{Constantin:2008aa},
\be\label{CIKelvin}
v_t= \mathbb{E} u_t, \qquad \oint_C v_t \cdot \rmd x = \mathbb{E}\oint_{A_t(C)} v_0 \cdot \rmd x,
\ee
in notation where $A_t= \Phi_t^{-1}$ and the map  $\Phi_t$ is the flow \eqref{dX-form} defined implicitly by the solution $v_t$.  The statement that \eqref{CIKelvin} holds for all rectifiable loops $C$ also characterizes the solution of the more general class of Lie-Laplacian Navier-Stokes (LLNS) equations in \eqref{LLNS-eqn}, as discussed in Drivas \& Holm  \cite{Drivas:2018aa}.  For further discussion of stochastic circulation and the Hamiltonian structure of the Navier-Stokes equation, see \cite{Drivas:2017aa,drivas2017lagrangian2}.

In summary,  the LA SALT Euler fluid equations in \eqref{LASALT-eqn} introduce a type of Lagrangian averaging, obtained by taking the expectation of the loop velocity $u^L_t$ in \eqref{dXhat-form} as the velocity of the Lagrangian parcels. The LA SALT loop velocity in the modified Kelvin theorem in \eqref{KelThm-Eul} will also be the transport velocity for advected quantities, when those quantities will be included in the dynamics. Thus, we may also derive LA SALT versions of compressible, adiabatic fluid dynamics, magnetohydrodynamics (MHD), etc. As mentioned earlier, taking the expectation  of the Lagrangian transport velocity in the Kelvin theorem for LA SALT is analogous to taking the temporal average at fixed  Lagrangian coordinate of the transport velocity for the Navier--Stokes equation to obtain the LANS-alpha equation \cite{Chen:1998aa,Chen:1999aa, Chen:1999ab, Foias:2001aa, Foias:2002aa}. The corresponding expected-quantity equations produce a Lie-Laplacian version of the Navier-Stokes equation.

\section{General case of LA SALT Theories}  \label{sec:LiePoisHam}

This section discusses the general theory of Lagrangian Averaged (LA) SALT, which is fundamentally based on Euler--Poincar\'e equations obtained from Hamilton's principle with symmetry-reduced Lagrangians and the Lie--Poisson Hamiltonian structures of these equations. We refer the reader to Appendix \ref{sec:GeomSet} for the definitions needed in the geometric setting for fluid dynamics. 

\vspace{1mm}\noindent\textbf{Probabilistic setting.} Let $(\Omega,\mathcal{F}, \mathbb{F}=\{\mathcal{F}_t\}_{t\ge 0}, \mathbb{P})$ be a filtered probability space satisfying the usual conditions of right-continuity and completeness. We assume the filtered probability space supports a sequence  of $\{W_t^{(k)}\}_{k\in \mathbb{N}, t\ge 0}$ of independent $\mathbb{R}$-valued $\mathbb{F}$-adapted Wiener process. Throughout the paper, the symbol $\xi =\{\xi^{(k)}\}_{k\in \mathbb{N}}\subset \mathfrak{X}$ will denote a collection of time-independent, deterministic, and divergence-free vector fields (i.e., $\operatorname{div}_g \xi^{(k)}=0$). 

We refer the reader to  \cite{revuz2013continuous} for more information about the basics of stochastic analysis and continuous martingale theory, to \cite{prevot2007concise} for more information about stochastic integration in Hilbert spaces and stochastic partial differential equations (SPDEs), and finally to  \cite{elliott2012stochastic, gyongy1993stochastic, gyongy1997stochastic} for more information about SPDEs on manifolds.

In sections \S \ref{sec:LiePoisHam},  \S \ref{sec:Vorticity} and \S \ref{sec:Helicity}, we will work on arbitrary Riemannian manifolds and assume whatever regularity and conditions are needed for the relations we calculate to hold properly. For the concrete example of the LA SALT Euler fluid equations in section \S \ref{Prob-sec}, however, we will provide a rigorous interpretation of the LA SALT equation as an SPDE on the torus.  In particular, we will specify the precise regularity required of $\xi$ and the other data in \S \ref{Prob-sec} to achieve wellposedness in a suitable sense.

\subsection{Euler--Poincar\'e and Lie--Poisson forms of the LA SALT equations}\label{sec:EPLP}
To take these introductory remarks further, we make the replacement \eqref{KelThm-form} in the Euler--Poincar\'e equation from which the SALT Kelvin circulation theorem arose in Stratonovich form, in \cite{Holm:2015aa}. That is, we keep the same physical class of reduced Lagrangians $\ell(u^L,a)$ as was treated for the SALT Euler--Poincar\'e variational principle in \cite{Holm:2015aa}.

Let $\ell=\ell(u^L,a): \mathfrak{X}\times V\rightarrow \mathbb{R}$ be the Lagrangian. We assume that  the Lagrangian possesses  Gateaux derivatives $\frac{\delta \ell}{\delta u^L}: \mathfrak{X}\times V\rightarrow \mathfrak{X}^*$ and $\frac{\delta \ell}{\delta a}: \mathfrak{X}\times V\rightarrow V^*$ defined by 
\begin{align}\label{def:GateuxLag}
\begin{split}
\lim_{\varepsilon\rightarrow 0} \ell(u+\varepsilon \delta u,a)&=\Scp{\frac{\delta \ell}{\delta u^L}}{\delta u}_{\mathfrak{X}}, \;\; \forall u,\delta u\in \mathfrak{X}, \; \forall a\in V\\
\lim_{\varepsilon\rightarrow 0} \ell(u,a+\varepsilon \delta a )&=\Scp{\frac{\delta \ell}{\delta a}}{\delta a}_{V}, \;\; \forall u\in \mathfrak{X}, \; \forall a,\delta a\in V.
\end{split}
\end{align} 
We also assume the Lagrangian is hyperregular; that is, the map $\delta \ell / \delta u^L(\cdot, a): \mathfrak{X}\rightarrow \mathfrak{X}^*$ is invertible  for every $a\in V$. The SALT equations for $\mathbb{F}$-adapted  $u^L: \Omega\times \mathbb{R}_+\rightarrow \mathfrak{X}$, $a: \Omega\times \mathbb{R}_+\rightarrow V$, and $\frac{\delta \ell}{\delta u^L}=\frac{\delta \ell}{\delta u^L}(u^L,a):\Omega\times \mathbb{R}_+\rightarrow \mathfrak{X}^*$, introduced in \cite{Holm:2015aa} read
\begin{align}\label{eq:SALT-EP}
\begin{split}
{\rm d}\frac{\delta \ell}{\delta u^L}+ \pounds_{ {\rm d}x_t} &\frac{\delta \ell}{\delta u^L}  
\overset{\mathfrak{X}^*}{=}  \frac{\delta \ell}{\delta a} \diamond a\,\rmd t
\quad\hbox{and}\quad
{\rm d}a + \pounds_{ {\rm d}x_t} a \overset{V}{=} 0,\\
\hbox{with} \quad \rmd x_t&= u^L_t \, \rmd t+ \sum_{k=1}^{\infty} \xi^{(k)}\circ \rmd W_t^{(k)}.
\end{split}
\end{align}
In this paper, we replace the stochastic transport vector field $ {\rm d}x_t$ in \eqref{eq:SALT-EP} with ${\rm d}X_t$ in \eqref{eq:LASALT-EP} to consider
\begin{align}\label{eq:LASALT-EP}
\begin{split}
{\rm d}\frac{\delta \ell}{\delta u^L}+ \pounds_{ {\rm d}X_t} &\frac{\delta \ell}{\delta u^L}
\overset{\mathfrak{X}^*}{=}  \E{ \frac{\delta \ell}{\delta a} }\diamond a\,\rmd t
\quad\hbox{and}\quad
{\rm d}a + \pounds_{ {\rm d}X_t} a \overset{V}{=} 0,\\
\hbox{with}\quad{\rm d}X_t&=\E{u^L_t}\, \rmd t + \sum_k\xi^{(k)}\circ dW_t^{(k)}.
\end{split}
\end{align}
The above equations comprise the class of LA SALT theories which we consider in the present paper.

Let  $D=\rho \,dV\in \operatorname{Den}$ denote  the mass density of the fluid, which is  one of the advected variables:
$$
dD+\pounds_{dX_t}D\overset{\Omega^d}{=}0.
$$
Upon using the product rule, we derive the Kelvin-Noether form (see Corollary 6.3 in \cite{Holm:1998aa}) of the LA SALT motion equation in \eqref{eq:LASALT-EP}, namely,
\begin{equation}\label{eq:KNf}
{\rm d}\left(\frac{1}{D}\frac{\delta \ell}{\delta u^L}\right)+ \pounds_{ {\rm d}X_t} \left(\frac{1}{D}\frac{\delta \ell}{\delta u^L}\right)
\overset{\Omega^1}{=} \frac{1}{D} \E{ \frac{\delta \ell}{\delta a} }\diamond a\,\rmd t
\,.\end{equation}
The  Kelvin-Noether Circulation theorem for LA SALT follows from the Kunita-It\^o-Wentzell formula  (see Theorem 3.3 of  \cite{Leon:2019aa}):
\begin{equation}\label{thm:KelvinThm}
\rmd\oint_{C(dX_t)}\frac{1}{D}\frac{\delta \ell}{\delta u^L} =\oint_{C(dX_t)}\frac{1}{D} \E{ \frac{\delta \ell}{\delta a} }\diamond a\,,
\end{equation}
where $C(dX_t)=\phi_t(C)$ is any closed-material loop $C$ in $M$ moving with the velocity $dX_t$.

Define the notation 
\begin{equation}\label{eq:upsilon}
\upsilon = \left(\frac{1}{D} \frac{\delta \ell}{\delta u^L} \right)^{\sharp}.
\end{equation}
Then  using the formula for $\operatorname{ad}^{\dagger}$ given in  \eqref{eq:addagger}, the condition  $\operatorname{div}_g \xi^{(k)}=0$, and the definition of $\hat{\diamond}$ given in \eqref{eq:hat_diamond}  we get
\begin{equation}\label{eq:KNsharp}
d\upsilon + \operatorname{ad}^{\dagger}_{dX_t}\upsilon + \operatorname{div}_g (\E{u_t^L})\upsilon \,  \rmd t\overset{\mathfrak{X}}{=} \E{ \frac{\delta \ell}{\delta a} }\hat{\diamond} \frac{a}{\rho} \, \rmd t.
\end{equation}

\noindent{\bf Legendre transform to the Hamiltonian side.} The Legendre transform from the Lagrangian side to the Hamiltonian side for SALT in  \cite{Holm:2015aa} is given by
\be
\mu=\frac{\delta \ell}{\delta u^L}\ \ \text{ and } \ \ \rmd h(\mu, a) = \scp{\mu}{{\rm d}x_t}_{\mathfrak{X}} - \ell(u^L,a) \rmd t,
\ee 
where ${\rm d}x_t$ is given in equation \eqref{dx-form1}. By introducing the deterministic Hamiltonian
\begin{equation}\label{biGh}
H(\mu,a):= \scp{\mu}{u^L}_{\mathfrak{X}} -  \ell(u^L,a),
\end{equation}
we may alternatively express the above as
\be
\rmd h(\mu,a) = H(\mu,a)dt + \sum_k \scp{\mu}{\xi^{(k)}}_{\mathfrak{X}} \circ \rmd W_t^{(k)}.
\ee

Note that $\delta h/\delta u^L=0$ and $\delta h/\delta a = \delta H/\delta a$ by definition. Taking variations of $h(\mu, a)$ yields
\begin{align}
\begin{split}
\delta h(\mu, a) &= \Scp{\delta \mu} {\frac{\delta h}{\delta \mu}}_{\mathfrak{X}}
+ \Scp{\frac{\delta h}{\delta a}}{\delta a}_{V}
\\&= \scp{\delta \mu}{{\rm d}x_t}_{\mathfrak{X}} 
- \Scp{\frac{\delta \ell}{\delta a}}{\delta a}_{V}
+ \Scp{\mu - \frac{\delta \ell}{\delta u^L}}{\delta u^L}_{\mathfrak{X}} 
\,.
\end{split}
\label{Leg-xform}
\end{align}
Then, upon identifying corresponding terms, one verifies the fibre derivative $\mu=\delta \ell/\delta u^L$ and finds the following variational derivatives of the SALT Hamiltonian, 
\begin{align}
\rmd\frac{\delta h}{\delta \mu} = {\rm d}x_t = { \frac{\delta H }{ \delta\mu} } {\rm d}t + \sum_k {\xi}^{(k)} \circ {\rm d}W_t^{(k)} 
\quad\hbox{and}\quad
{\frac{\delta H}{\delta a}} = -\,{\frac{\delta \ell}{\delta a}}
\,.\label{h-vars-SALT}
\end{align}
At this point, we take expectations of the terms in \eqref{h-vars-SALT} which pass from the SALT equations in \eqref{eq:SALT-EP} to the LA SALT equations in \eqref{eq:LASALT-EP}. Specifically, defining for LA SALT
\begin{align*}
\mu=\frac{\delta \ell}{\delta u^L}\ \ \text{ and } \ \ \rmd h(\mu, a) &= \scp{\mu}{{\rm d}X_t}_{\mathfrak{X}} - {\ell(u^L,a)} \rmd t
\end{align*}
we obtain the Hamiltonian formulation of the LA SALT equations from the Lagrangian formulation above 
\begin{align}
\rmd \frac{\delta h}{\delta \mu} = {\rm d}X_t = \E{ \frac{\delta H }{ \delta\mu} }\,  {\rm d}t + \sum_k {\xi}^{(k)} \circ {\rm d}W_t^{(k)} 
\quad\hbox{and}\quad
\E{\frac{\delta H}{\delta a}} = -\,\E{\frac{\delta \ell}{\delta a}}
\,,\label{h-vars-LASALT}
\end{align}
where $H$ is again defined by \eqref{biGh}.
Taking the expectation then transforms the LA SALT equations \eqref{eq:LASALT-EP} in Euler--Poincar\'e into Hamiltonian form with a  Lie--Poisson matrix operator. 
Thus, the Stratonovich version of the SDP-LPB for the SALT Hamiltonian formulation yields the LA SALT equations in \eqref{eq:LASALT-EP} as
\begin{align}\label{eq:SDP-LASALT-Strat}
{\rm d} 
\begin{bmatrix}
\mu 
\\ \\
a
\end{bmatrix}
= -
\begin{bmatrix}
\pounds_{(\,\cdot\,)}\mu   & (\,\cdot\,)\diamond a
\\ \\
\pounds_{(\,\cdot\,)} a          & 0
\end{bmatrix}
\begin{bmatrix}
\E{ \delta H / \delta\mu } \, {\rm d}t + \sum_k {\xi}^{(k)} \circ {\rm d}W_t^{(k)} 
\\ \\
\E{ \delta H/ \delta a}\,  {\rm d}t 
\end{bmatrix}
.\end{align}
The definition of the diamond operator $(\diamond)$ \eqref{diamond-def} ensures that the Lie--Poisson matrix operator is skew-symmetric under integration by parts in $L^2$ pairing. 
The modification of the Hamiltonian form of the SALT equations to obtain the LA SALT equations in \eqref{eq:SDP-LASALT-Strat} replaces the variational derivative of the Hamiltonian with respect to momentum and to the advected variable by their expected values. This modification \emph{preserves} the Hamiltonian matrix operator in both the deterministic and SALT formulations and makes that operator available for exploring the solution behaviour for LA SALT, as we discuss below in a combination of theorems and illustrative examples.  Essentially, we will discover below that the LA SALT theory is a nonequilibrium stochastic linear response theory for fluctuations in SALT fluids with advected quantities.

\begin{remark} We will show below that the expectation of the system \eqref{eq:SDP-LASALT-Strat} results in a closed dynamical system, when the expected variational derivatives ${ \delta H / \delta\mu }$ and ${ \delta H / \delta a }$ are linear in the  variables ${\mu}$ and ${a}$, after one has accounted for the constraints (e.g. incompressibility). As we shall see, upon regarding SALT as the `mother theory', LA SALT can be regarded as a first-order cumulant discard closure for SALT and therefore it can be characterized as a type of linear response theory, particularly because its dynamics involves both fluctuations and dissipation. We will investigate several examples of this situation in the remainder of the paper. 
\end{remark}

\noindent{\bf Casimirs.} The LA SALT system \eqref{eq:SDP-LASALT-Strat} undergoes \emph{stochastic coadjoint motion}. That is, for any given functional $C: \mathfrak{X}^*\times V^*\rightarrow \mathbb{R}$, the LA SALT dynamics of $C[\mu,a]$ is given by
\begin{align}\label{eq:SDP-LASALT-Strat-Cas}
\begin{split}
{\rm d}  C[\mu,a]&=\Scp{d\mu}{\frac{\delta C}{\delta \mu}}_{\mathfrak{X}} + \Scp{\frac{\delta C}{\delta a}}{da}_{V}=\Scp{d\mu}{\frac{\delta C}{\delta \mu}}_{\mathfrak{X}} + \Scp{da}{\frac{\delta C}{\delta a}}_{V^*}\\
&= -\left \langle
\begin{bmatrix}
\pounds_{(\,\cdot\,)}\mu   & (\,\cdot\,)\diamond a
\\ \\
\pounds_{(\,\cdot\,)} a          & 0
\end{bmatrix}
\begin{bmatrix}
\E{ \delta H / \delta\mu }\,  {\rm d}t + \sum_k {\xi}^{(k)} \circ {\rm d}W_t^{(k)} 
\\ \\
\E{ \delta H/ \delta a} \, {\rm d}t 
\end{bmatrix},
\begin{bmatrix}
\delta C / \delta\mu 
\\ \\
\delta C / \delta a 
\end{bmatrix}\right\rangle_{\mathfrak{X}\oplus V^*}.
\end{split}
\end{align}
A functional $C[\mu,a]$ whose variational derivatives $[\delta C / \delta\mu ,\delta C / \delta a ]^T$ comprise a null eigenvector of the Hamiltonian matrix operator in \eqref{eq:SDP-LASALT-Strat-Cas} is called a \emph{Casimir functional} for that Lie--Poisson system. Casimir functionals satisfy ${\rm d} C[\mu,a]=0$, so that $C[\mu_t,a_t]=C[\mu_0,a_0]$ \emph{for any Hamiltonian $H[\mu,a]$}. By having preserved the Lie--Poisson structure of the deterministic Hamiltonian fluid equations in formulating the LA SALT system \eqref{eq:SDP-LASALT-Strat}, one has preserved the Casimir conserved quantities for the original deterministic Lie--Poisson fluid dynamics. In turn, one has also preserved the expectations of the Casimirs, since for them  $\E{C[\mu_t,a_t]}=\E{C[\mu_0,a_0]}$.

Thus, equation \eqref{eq:SDP-LASALT-Strat-Cas} for LA SALT encapsulates all three of the stages in the approximations of fluid dynamics which we have been discussing. The SALT formulation in \cite{Holm:2015aa} emerges when the expectations are not taken in the first term of the product in the integrand. The historical deterministic formulation discussed in the Introduction emerges when the ${\xi}^{(k)}$ also vanish in that term. Since the Casimirs are defined as null vectors of the same Hamiltonian operator in each case, they persist in all three stages of deterministic, SALT and LA SALT fluid dynamics. 

In summary, because the LA SALT modification of the SALT transport vector field preserves the form of the reduced Euler--Poincar\'e Lagrangian in \eqref{eq:LASALT-EP} and the Lie--Poisson Hamiltonian operator in \eqref{eq:SDP-LASALT-Strat}, one retains both the Kelvin circulation theorem and the conservation of Casimirs of the Lie--Poisson bracket in the LA SALT dynamics. 

\subsection{Closed dynamics of the average}
Significant simplifications occur when the drift velocity of SALT is replaced by its expectation in LA SALT.  Indeed, upon converting \eqref{eq:LASALT-EP} to It\^o-form, we find
\begin{align}\label{eq:LASALT-EP-Ito}
\begin{split}
{\rm d}\mu+ \pounds_{ \E{\frac{\delta H}{ \delta\mu}}} \mu \, \rmd  t   + \pounds_{\xi^{(k)}}\mu  \, \rmd W_t^{(k)} 
&=\left( \frac{1}{2}\sum_k  \pounds_{\xi^{(k)}}( \pounds_{\xi^{(k)}}\mu) \,  \rmd t-\E{ \frac{\delta H}{\delta a}} \diamond a\right)\,\rmd t\\
{\rm d}a + \pounds_{ \E{\frac{\delta H}{ \delta\mu}}} a \, \rmd t   + \pounds_{\xi^{(k)}}a \,  \rmd W_t^{(k)}    & = \frac{1}{2}\sum_k  \pounds_{\xi^{(k)}}( \pounds_{\xi^{(k)}}a)\, \rmd t
\,.
\end{split}
\end{align}
Taking the expectation of \eqref{eq:LASALT-EP-Ito} then yields
\begin{align}\label{eq:LASALT-EP-Exp}
\begin{split}
\frac{\rmd }{\rmd t} \E{\mu} + \pounds_{ \E{\frac{\delta H}{\delta\mu}}} \E{\mu} 
&=\frac{1}{2}  \sum_k \pounds_{\xi^{(k)}}( \pounds_{\xi^{(k)}} \E{\nu})   - \,\E{ \frac{\delta H}{\delta a}}\diamond \E{a}
\,,\\
\frac{\rmd }{\rmd t} \E{a} +  \pounds_{ \E{\frac{\delta H}{\delta\mu}}} \E{a} &= \frac{1}{2}  \sum_k \pounds_{\xi^{(k)}}( \pounds_{\xi^{(k)}} \E{a})
\,.
\end{split}
\end{align}
Similarly, starting from \eqref{eq:KNsharp}    we obtain
\begin{equation}\label{eq:expKNf}
\partial_t \E{\upsilon } + \operatorname{ad}^{\dagger}_{ \E{\frac{\delta H}{\delta\mu}}}\E{\upsilon} + \operatorname{div}_g \left(\E{\frac{\delta H}{\delta\mu}}\right)\E{\upsilon}
=\frac{1}{2}  \sum_k \operatorname{ad}^{\dagger}_{\xi^{(k)}}( \operatorname{ad}^{\dagger}_{\xi^{(k)}} \E{\upsilon}) - \,\E{ \frac{\delta H}{\delta a}}\hat{\diamond}\, \E{\frac{a}{\rho}},
\end{equation}
where $\hat{\diamond}$ is defined by \eqref{eq:hat_diamond}.
These equations provide the history of the expectations $\E{\mu}$, $\E{\upsilon}$, and $\E{a}$ throughout the duration of the flow. If, for example, the variations $\delta H/\delta \mu$ and $\delta H/\delta a$ are linear in $\mu$ and $a$, then the equations  \eqref{eq:LASALT-EP} (equivalently, \eqref{eq:LASALT-EP-Ito}) are  slaved to the  expectations $\E{\mu}$ and $\E{a}$  as linear stochastic transport relations. We refer the reader to the numerous examples in Section  \ref{sec:examples} for which this is the case, particularly when the flow is divergence free.

Thus, introduction of nonlocality in the sense of McKean  \cite{McKean:1966aa}  in the LA SALT equations \eqref{eq:LASALT-EP}  has significantly simplified stochastic fluid dynamics in two ways. First, it preserves the differential structure and form of the nonlinear deterministic fluid motion and advection equations that result in promotion of Lagrangian conservation laws to our setting. Second, it introduces linear equations for the fluctuations (in many special cases, including incompressible Euler), which are stochastically driven while being transported by the expectation velocity and accelerated by forces involving expectations.

\subsection{Fluctuation variance dynamics} \label{sec:FluctDyn}
In this section, we will discuss the fluctuations of \eqref{eq:LASALT-EP}  and \eqref{eq:KNsharp} about their average. We first define the fluctuation variables by
\begin{align*}
\upsilon' := \upsilon - \E{\upsilon} \in \mathfrak{X}, \quad  a' := a - \mathbb E[a] \in V
\end{align*}
The dynamics of the fluctuations are obtained by taking the difference between the Stratonovich formulations \eqref{eq:LASALT-EP} and \eqref{eq:LASALT-EP-Exp}
\begin{align}\label{eq:LASALT-EP-Fluc-Strat}
\begin{split}
{\rm d}\upsilon'&+ \left[\operatorname{ad}^{\dagger}_{ \E{\frac{\delta H}{ \delta\mu}}} \upsilon'+\operatorname{div}_g \left(\E{\frac{\delta H}{\delta\mu}}\right)\upsilon'\right]\rmd  t  + \pounds_{\xi^{(k)}}\upsilon  \circ  \rmd W_t^{(k)} \\
&=\left[- \frac{1}{2}\sum_k  \operatorname{ad}^{\dagger}_{\xi^{(k)}}( \operatorname{ad}^{\dagger}_{\xi^{(k)}}\E{\upsilon}) - \E{ \frac{\delta H}{\delta a}} \tilde{\diamond}\, \left(\frac{a}{\rho}\right)'\right]\,\rmd t,\\
{\rm d}a' &+ \pounds_{ \E{\frac{\delta H}{ \delta\mu}}} a' \, \rmd t   + \pounds_{\xi^{(k)}}a \circ   \rmd W_t^{(k)}     = - \frac{1}{2}\sum_k  \pounds_{\xi^{(k)}}( \pounds_{\xi^{(k)}}\E{a})\,  \rmd t\,,
\end{split}
\end{align}
where $(\frac{a}{\rho})'=\frac{a}{\rho}-\E{\frac{a}{\rho}}$.
The corresponding It\^o-formulation of the  fluctuation dynamics, obtained, for example, by taking the difference of \eqref{eq:LASALT-EP-Ito} and \eqref{eq:LASALT-EP-Exp}, are
\begin{align}\label{eq:LASALT-EP-Fluc-Ito}
\begin{split}
{\rm d}\upsilon'&+  \left[\operatorname{ad}^{\dagger}_{ \E{\frac{\delta H}{ \delta\mu}}} \upsilon'+\operatorname{div}_g \left(\E{\frac{\delta H}{\delta\mu}}\right)\upsilon'\right]\, \rmd t  + \operatorname{ad}^{\dagger}_{\xi^{(k)}}\upsilon  \,\rmd W_t^{(k)} \\
&=\left[\frac{1}{2}\sum_k  \operatorname{ad}^{\dagger}_{\xi^{(k)}}( \operatorname{ad}^{\dagger}_{\xi^{(k)}}\upsilon') - \E{ \frac{\delta H}{\delta a}} \tilde{\diamond}\, \left(\frac{a}{\rho}\right)'\right]\,\rmd t,\\
{\rm d}a' &+ \pounds_{ \E{\frac{\delta H}{ \delta\mu}}} a' \, \rmd t   + \pounds_{\xi^{(k)}}a \, \rmd W_t^{(k)}  = \frac{1}{2}\sum_k  \pounds_{\xi^{(k)}}( \pounds_{\xi^{(k)}}a')\,  \rmd t\,.
\end{split}
\end{align}
This formulation simplifies the calculations in this section. Applying the  It\^o-product rule pointwise, we find
\begin{align}\label{eq:LASALT-EP-Fluc-point}
\begin{split}
\frac12{\rm d}g(\upsilon',\,\upsilon')&+\left[g\left(\operatorname{ad}^{\dagger}_{ \E{\frac{\delta H}{ \delta\mu }}} \upsilon'+\operatorname{div}_g \left(\E{\frac{\delta H}{\delta\mu}}\right)\upsilon',\,\upsilon'\right) \right] \, \rmd  t   +\sum_k g\left(\operatorname{ad}^{\dagger}_{\xi^{(k)}}\upsilon,\,\upsilon'\right)\, \rmd W_t^{(k)}\\
&= \frac12\sum_k  \left[g\left(\operatorname{ad}^{\dagger}_{\xi^{(k)}}( \operatorname{ad}^{\dagger}_{\xi^{(k)}}\upsilon')-\E{ \frac{\delta H}{\delta a}} \tilde{\diamond} \,\left(\frac{a}{\rho}\right)',\,\upsilon'\right) + g\left(\operatorname{ad}^{\dagger}_{\xi^{(k)}}\upsilon, \,\operatorname{ad}^{\dagger}_{\xi^{(k)}}\upsilon\right)\right]\,\rmd t,\\
\frac12{\rm d}g(a', \,a')&+ g\left(\pounds_{\E{\frac{\delta H}{ \delta\mu}} }a', \,a'\right)\,  \rmd  t   +\sum_k g\left(\pounds_{\xi^{(k)}}a,\, a'\right)\, \rmd W_t^{(k)}\\
&=\frac12 \sum_k  \left[g\left(\pounds_{\xi^{(k)}}( \pounds_{\xi^{(k)}}a'),\, a'\right)  +  g\left(\pounds_{\xi^{(k)}}a, \, \pounds_{\xi^{(k)}}a\right)\right]\,\rmd t.
\end{split}
\end{align}
Integrating \eqref{eq:LASALT-EP-Fluc-point} and using  stochastic Fubini's theorem (see, e.g., \cite{Krylov:2010aa}) yields
\begin{align}\label{eq:LASALT-EP-L2}
\begin{split}
\frac12{\rm d}|\upsilon'|^2_{L^2}&+\left[\left(\operatorname{ad}^{\dagger}_{ \E{\frac{\delta H}{ \delta\mu }}} \upsilon'+\operatorname{div}_g \left(\E{\frac{\delta H}{\delta\mu}}\right)\upsilon',\,\upsilon'\right)_{L^2}  \right]\, \rmd  t   +\sum_k \left(\operatorname{ad}^{\dagger}_{\xi^{(k)}}\upsilon,\,\upsilon'\right)_{L^2}\, \rmd W_t^{(k)}\\
&= \frac12\sum_k  \left[\left(\operatorname{ad}^{\dagger}_{\xi^{(k)}}( \operatorname{ad}^{\dagger}_{\xi^{(k)}}\upsilon')-\E{ \frac{\delta H}{\delta a}} \tilde{\diamond} \,\left(\frac{a}{\rho}\right)',\,\upsilon'\right)_{L^2} + \left(\operatorname{ad}^{\dagger}_{\xi^{(k)}}\upsilon, \,\operatorname{ad}^{\dagger}_{\xi^{(k)}}\upsilon\right)_{L^2}\right]\,\rmd t,\\
\frac12{\rm d}|a'|_{L^2}&+ \left(\pounds_{\E{\frac{\delta H}{ \delta\mu}} }a', \,a'\right)_{L^2}\,  \rmd  t   +\sum_k \left(\pounds_{\xi^{(k)}}a,\, a'\right)_{L^2}\, \rmd W_t^{(k)}\\
&=\frac12 \sum_k  \left[\left(\pounds_{\xi^{(k)}}( \pounds_{\xi^{(k)}}a'),\, a'\right)_{L^2}  +  \left(\pounds_{\xi^{(k)}}a, \, \pounds_{\xi^{(k)}}a\right)_{L^2}\right]\,\rmd t.
\end{split}
\end{align}
We remark that one could have directly deduced \eqref{eq:LASALT-EP-L2} from \eqref{eq:LASALT-EP-Fluc-Ito} under weaker assumptions than classical solutions by appealing to It\^o's formula for the square of the norm (see,  \cite{Pardoux:1972aa, Pardoux:1975aa, Krylov:1979aa, Krylov:2013aa,Krylov:2010aa}). Upon making use of equations \eqref{eq:addaggerrel}, \eqref{eq:hat_diamond}, \eqref{eq:advectedsym},  and \eqref{diamond-def}, we find
\begin{align}\label{eq:LASALT-EP-Fluc-int2}
\begin{split}
\frac12{\rm d}|\upsilon'|_{L^2}^2&-\left[\left(\operatorname{ad}^{\dagger}_{\upsilon'} \upsilon',\,\E{\frac{\delta H}{ \delta\mu }}\right)_{L^2}  +\SCP{\E{\frac{\delta H}{ \delta a }}}{\pounds_{ \upsilon} \left(\frac{a}{\rho}\right)'}_V \right]\, \rmd  t   \\
&=- \frac12\sum_k  \left(\left[\operatorname{ad}^{\dagger}_{\upsilon'}( \operatorname{ad}^{\dagger}_{\xi^{(k)}}\upsilon')+\operatorname{ad}_{\upsilon}( \operatorname{ad}^{\dagger}_{\xi^{(k)}}\upsilon) \right], \, \xi^{(k)}\right)_{L^2}\,\rmd t \\
&\quad- \left(\operatorname{div}_g \left(\E{\frac{\delta H}{\delta\mu}}\right)\upsilon',\,\upsilon'\right)_{L^2}+\sum_k \left(\operatorname{ad}^{\dagger}_{\upsilon}\upsilon,\,\xi^{(k)}\right)_{L^2}\, \rmd W_t^{(k)},\\
\frac12{\rm d}|a'|_{L^2}^2&- \SCP{\star a'\diamond a'}{\E{\frac{\delta H}{ \delta\mu}}}_{\mathfrak{X}}  \rmd  t   -\sum_k \SCP{ \star a'\diamond a}{\xi^{(k)}}_{\mathfrak{X}}\, \rmd W_t^{(k)}\\
&=-\frac12 \sum_k  \SCP{\left[\star a'\diamond ( \pounds_{\xi^{(k)}}a')+\star \pounds_{\xi^{(k)}}a\diamond a\right]}{\xi^{(k)}}_{\mathfrak{X}}  \,\rmd t.
\end{split}
\end{align}

Finally, upon taking the expectation of \eqref{eq:LASALT-EP-Fluc-int2}, we find 
\begin{align}\label{eq:LASALT-EP-var}
\begin{split}
\frac12\partial_t\E{|\upsilon'|_{L^2}^2}&- \left(\E{\operatorname{ad}^{\dagger}_{\upsilon'} \upsilon'}, \, \E{\frac{\delta H}{ \delta\mu}}\right)_{L^2}- \SCP{\E{ \frac{\delta H}{\delta a}}}{\E{ \pounds_{\upsilon}\left(\frac{a}{\rho}\right)'}}_{V} \\
&=- \frac12\sum_k  \left(\bE{\left[\operatorname{ad}^{\dagger}_{\upsilon'}( \operatorname{ad}^{\dagger}_{\xi^{(k)}}\upsilon')+\operatorname{ad}_{\upsilon}( \operatorname{ad}^{\dagger}_{\xi^{(k)}}\upsilon) \right]}, \, \xi^{(k)}\right)_{L^2}\\
&\quad -\left(\E{g(\upsilon',\,\upsilon')},\operatorname{div}_g \left(\E{\frac{\delta H}{\delta\mu}}\right)\right)_{L^2}\,, \\
\frac12\partial_t\E{|a'|_{L^2}^2}&- \SCP{\E{\star a'\diamond a'}}{\E{\frac{\delta H}{ \delta\mu}}}_{\mathfrak{X}} =-\frac12 \sum_k  \SCP{\E{\star a'\diamond ( \pounds_{\xi^{(k)}}a')+\star \pounds_{\xi^{(k)}}a\diamond a}}{\xi^{(k)}}_{\mathfrak{X}} \,.
\end{split}
\end{align}

In particular, upon assuming  $\operatorname{div}_g \E{\frac{\delta H}{\delta \mu}}=0$ and $\rho\equiv 1$ we obtain
\begin{align}\label{eq:daggerfluc}
\frac12\partial_t\E{|\upsilon' |_{L^2}^2}&- \left(\E{\operatorname{ad}^{\dagger}_{\upsilon'} \upsilon'}, \, \E{\frac{\delta H}{\delta \mu}}\right)_{L^2}  - \SCP{\E{ \frac{\delta H}{\delta a}}}{\E{ \pounds_{\upsilon }a'}}_{V}\\
&=- \frac12\sum_k  \left(\bE{\left[\operatorname{ad}^{\dagger}_{\upsilon }( \operatorname{ad}^{\dagger}_{\xi^{(k)}}\upsilon')+\operatorname{ad}_{\upsilon}( \operatorname{ad}^{\dagger}_{\xi^{(k)}}\upsilon\right]}, \, \xi^{(k)}\right)_{L^2}.
\end{align}
Thus, we can see that the  dynamics of the variances of the stochastic system \eqref{eq:LASALT-EP-Fluc-Strat} is driven by an intricate variety of correlations among the evolving fluctuation variables.   The consequences of these very general equations can be seen more easily in examples.  A special case will be the vorticity dynamics in the LA SALT Euler fluid equations we shall treat next. 

\section{An illustrative example: LA SALT Euler.}\label{sec:LASALT-Eu}
We now show that the LA SALT Euler equations introduced in the beginning of the paper comprise a special case of the general class of the Lie--Poisson Hamiltonian systems defined by \eqref{eq:LASALT-EP}.

For a given  $p\in \Omega^0$ on $M$, let us define  $\ell : \mathfrak{X}\times V\rightarrow \mathbb{R}$ for all $u\in \mathfrak{X}$ and $D=\rho dV\in V=\operatorname{Den}$ by
$$\ell(u,D)=\int_{M}\left(\frac{\rho}{2}g(u,u) - p(\rho-1)\right) dV=\int_{M}\left(\frac{\rho}{2}u^{\flat}(u)  - p(\rho-1)\right)dV.$$
Let $V^*= \Omega^0$ and define
\begin{equation*}
\langle b, D \rangle_{V}:=\int_{M}b\rho dV,\qquad D=\rho dV\in V, \;\;b\in V^*.
\end{equation*}
Then
$$ 
\mu=\frac{\delta \ell}{\delta u}  = u^{\flat}\otimes D\in \mathfrak{X}^*\quad \textnormal{and} \quad \frac{\delta \ell}{\delta D}=-\frac{\delta H}{\delta D}=\frac{1}{2}u^{\flat}(u)dt - p\in V^*,
$$
and $\delta H/ \delta \mu = u$. Using the Leibniz property of the  Lie derivative, Stoke's theorem (with $\partial M=\emptyset$) yields
\begin{align}\label{eq:diamondcalc}
\begin{split}
\langle b, -\pounds_u D\rangle_V &= -\int_{M} b \pounds_u (\rho dV)=-\int_{M}\left[ \pounds_{u}(b\rho dV) - (\pounds_{u}b) \rho dV\right] \\
&= -\int_{M} \mathbf{d}\mathbf{i}_u(b\rho dV) +\int_{M} (\mathbf{i}_u\mathbf{d}b) \rho dV\\
&=\langle \mathbf{d}b \otimes  D, u\rangle_{\mathfrak{X}}, \;\; \forall D=\rho dV\in V, \;\; \forall b\in V^*,\;\;  \forall u\in \mathfrak{X}.
\end{split}
\end{align}
which implies that 
$
b\diamond D=-\mathbf{d}b \otimes  D\in \mathfrak{X}^*.
$
It follows that  
$$
\E{\frac{\delta \ell}{\delta D}}\diamond D=\E{\frac{1}{2}\mathbf{d}\left(u^{\flat}(u)\right)-\mathbf{d}p}\otimes D.
$$
The LA SALT equations are obtained by substituting this relation into   \eqref{eq:LASALT-EP}, which in this case read
$$
{\rm d}\left(u^{\flat}\otimes D \right)+ \pounds_{ {\rm d}X_t} \left(u^{\flat}\otimes D\right)
=\E{\frac{1}{2}\mathbf{d}\left(u^{\flat}(u)\right)-\mathbf{d}p}\otimes D\,\rmd t
\quad\hbox{and}\quad
{\rm d}D + \pounds_{ {\rm d}X_t} D = 0.
$$
Recalling $D=\rho dV \in V$  and   that the Lie derivative satisfies the Leibniz rule with respect to the tensor product, we have
$$
{\rm d}u^{\flat}+ \pounds_{ {\rm d}X_t} u^{\flat}
=\mathbf{d}\left(\E{\frac{1}{2}u^{\flat}(u)-p}\right)\rmd t
\quad\hbox{and}\quad
{\rm d}\rho + \pounds_{ {\rm d}X_t} \rho  + \operatorname{div}({\rm d}X_t)\rho = 0.
$$
We now discuss how the `pressure' is determined. We understand `incompressibility' in this example to mean that the  density $D=\rho dV$, which is Lie-advected by $dX_t$, satisfies $\rho\equiv 1$. 
Since the density $\rho$ satisfies 
$${\rm d}\rho + \pounds_{ {\rm d}X_t} \rho =- \operatorname{div}({\rm d}X_t)\rho,$$
the only term driving the density away from unity is 
\begin{equation}
\operatorname{div}({\rm d}X_t)= \operatorname{div}(\E{u_t})\rmd t+\operatorname{div}(\xi^{(k)}) \rmd W_t^{(k)}= \operatorname{div}(\E{u_t})\rmd t,
\end{equation}
where we have used that $\operatorname{div}(\xi^{(k)}) =0$ by assumption. Therefore, to impose that $\rho \equiv 1$, we require $\operatorname{div}(\E{u_t})=  0$. Upon setting $\tilde{p}:=\frac{1}{2}u^{\flat}(u)-p$, we arrive at the LA SALT Euler equations  on $M$ 
\begin{align*}
{\rm d}u^{\flat}+ \pounds_{ \E{u}} u^{\flat}\rmd t +\sum_k \pounds_{\xi^{(k)}} u^{\flat}\circ \rmd W_t^{(k)}
&=\mathbf{d}\E{\tilde{p}}\,\rmd t\\
\delta \E{u^{\flat}}&=0,
\end{align*}
where  $\delta: \Omega^1\rightarrow \Omega^0$ is the codifferential operator, and where $\tilde{p}$ is non-locally determined to enforce $\delta \E{u^{\flat}}=0$. Recalling that $\mathbf{P}$ is the divergence-free projection operator and letting $\mathbf{P}^{\mathbb{E}}=I - \mathbf{Q}\mathbb{E}$, we can then rewrite the equations as
\begin{equation}\label{eq:LASALT-Euler-Proj}
{\rm d}u^{\flat}+ \mathbf{P}^{\mathbb{E}}\pounds_{ \E{u}} u^{\flat}\, \rmd t + \sum_k\mathbf{P}^{\mathbb{E}}\pounds_{\xi^{(k)}} u^{\flat}\circ \rmd W_t^{(k)}=0.
\end{equation}
We note that this equation is equivalent to  \eqref{LASALT-eqn} in the Introduction, where $u^{\flat}= u\cdot dx$. Thus, it clear that the pressure required to maintain incompressibility of  $\E{u_t}$ is deterministic. The  reason for treating the pressure to be $\mathbb{F}$-adapted a priori rather than deterministic is to maintain the connection with the structure introduced in the general class of LA SALT models \eqref{eq:SDP-LASALT-Strat} introduced in Section \ref{sec:EPLP}.  However, given the dynamics, only the expectation of the pressure can be recovered, and hence it is only the expected pressure which plays a role.  Thus, hereafter, we will simply denote $ \pi_t =  \E{p}$ and write
\begin{align}\label{eq:LASALT-Euler}
\begin{split}
{\rm d}u^{\flat}+ \pounds_{ \E{u}} u^{\flat}\, \rmd t +\sum_k \pounds_{\xi^{(k)}} u^{\flat}\circ \rmd W_t^{(k)}
&=\mathbf{d}\pi\,\rmd t\\
\delta \E{u^{\flat}}&=0.
\end{split}
\end{align}
\begin{remark}
Instead of defining incompressibility to mean  that $\rho \equiv 1$, we could have defined it to mean that the  stochastic velocity $u=(u^\flat)^\sharp=\upsilon$ is divergence free. While we do not pursue this here, it can easily be incorporated into our framework by defining $\mathfrak{X}=\mathfrak{X}_{\operatorname{div}}$ and $\mathfrak{X}^*=\Omega^1/ \mathbf{d} \Omega^0$ and letting 
$$
\ell(u)=\int_{M}g(u,u)dV=\int_{M}u^{\flat}(u)dV.
$$
In this case, the LA SALT Euler motion  equation reads, in the notation of Appendix \ref{sec:GeomSet},
\begin{equation}
{\rm d}u^{\flat}+ \mathbf{P}\pounds_{ \E{u}} u^{\flat}\, \rmd t + \sum_k\mathbf{P}\pounds_{\xi^{(k)}} u^{\flat}\circ \rmd W_t^{(k)}=0,
\end{equation}
or, equivalently, upon substituting $\mathbf{P}= I - \mathbf{Q}$, we have
\begin{align*}
{\rm d}u^{\flat}+ \pounds_{ \E{u}} u^{\flat}\, \rmd t + \sum_k\pounds_{\xi^{(k)}} u^{\flat}\circ \rmd W_t^{(k)}=\mathbf{d} p\,  \rmd t +\sum_k\mathbf{d} q^{(k)} \rmd W_t^{(k)}
\quad\hbox{and}\quad
\delta u^{\flat}&=0,
\end{align*}
where
$$
\mathbf{d} p=\mathbf{Q}\pounds_{ \E{u}} u^{\flat}+\mathbf{Q}\sum_k (\pounds_{\xi^{(k)}})^2u^{\flat}
\quad\hbox{and}\quad
  \mathbf{d} q^k=\mathbf{Q}\pounds_{\xi^{(k)}} u^{\flat}.
$$
We remark that the analytical results stated in Section \ref{Prob-sec} for the model \eqref{eq:LASALT-Euler}  on $M=\mathbb{T}^d$ are also true for this model. 
\end{remark}

\subsection{Vorticity dynamics}\label{sec:Vorticity}

Let $\omega= du^{\flat}\in \Omega^2$ denote the vorticity two-form. Applying the exterior derivative $\mathbf{d}$ to \eqref{eq:LASALT-Euler} and using the property that it commutes with the Lie derivative yields
\begin{equation}\label{eq:LASALT-Euler-Vort}
{\rm d}\omega+ \pounds_{\rmd X_t} \omega={\rm d}\omega+ \pounds_{ \E{u}} \omega\, \rmd t +\sum_k \pounds_{\xi^{(k)}} \omega  \circ \rmd W_t^{(k)}=0.
\end{equation}
In computing the fluctuations of $\omega$, we may treat $\omega$ as an advected variable $a\in \Omega^2$, and thus from  \eqref{eq:LASALT-EP-Fluc-int2}, we find
$$
\frac12\partial_t\E{|\omega'|_{L^2}^2}- \SCP{\E{\star \omega'\diamond \omega' }}{\E{u}}_{\mathfrak{X}} =-\frac12 \sum_k  \SCP{\E{\star \omega '\diamond ( \pounds_{\xi^{(k)}}\omega')+\star \pounds_{\xi^{(k)}}\omega\diamond \omega}}{\xi^{(k)}}_{\mathfrak{X}}.
$$ 
Therefore, the correlates $\xi^{(k)}$  play a decisive role in balancing the spatially integrated variances of the  fluctuations of the  vorticity.

\vspace{1mm}\noindent\textbf{Three dimensions.}
In 3D, the vorticity $\omega$ may be identified with the vector  $\vec{\omega}=\sharp \star \omega \in \mathfrak{X}$. Using the identity $[\sharp\star, \pounds_v]=0$ (see, e.g., appendix section A.6. of 
\cite{Besse:2017aa}), we find that $\vec{\omega}$  is governed by the following system:
\begin{equation}
{\rm d}\vec{\omega}+ \pounds_{\rmd X_t} \vec{\omega}= {\rm d}\vec{\omega}+ [\rmd X_t, \vec{\omega}] ={\rm d}\vec{\omega}- \operatorname{ad}_{\rmd X_t}\vec{\omega}={\rm d}\vec{\omega}- \operatorname{ad}_{\E{u}}\vec{\omega}\, \rmd t -\frac{1}{2}\sum_k\operatorname{ad}_{\xi^{(k)}}\vec{\omega}\circ \rmd W^{(k)}_t\, =0.
\end{equation}
Following the analysis of Section \ref{sec:FluctDyn} and using the definition  \eqref{eq:addaggerrel}, we find 
$$
\frac{1}{2}\frac{\rmd}{\rmd t}\mathbb{E}\left[|\vec{\omega}'|_{L^2}^2\right]+\left(\mathbb{E}\left[{\rm ad}^{\dagger}_{\vec{\omega}_t' }\vec{\omega}_t'\right],\,   \E{u_t} \right)_{L^2}+\frac{1}{2}\sum_{k} \left(\mathbb{E}\left[{\rm ad}_{\vec{\omega}_t' }^{\dagger}{\rm ad}^{\dagger}_{\xi^{(k)}}\vec{\omega}_t'+{\rm ad}_{\vec{\omega}_t }^{\dagger}{\rm ad}_{ \xi^{(k)}} \vec{\omega}_t\right], \xi^{(k)}\right)_{L^2}  =0.
$$
Although the three-dimensional vorticity dynamics discussed so far is geometrically meaningful, it is not particularly illuminating for understanding the qualitative behaviour of the properties of its vorticity solutions.  In two dimensions, we can obtain much more information about the vorticity solution behaviour.

 \vspace{1mm}\noindent\textbf{Two dimensions.} 
The vorticity  in 2D  may be transformed to a scalar using the Hodge-star operator $\omega = \star \omega\in \Omega^0$. (We abuse notation and still call it $\omega$ for the purposes of this section.) Again, using the identity $[\star, \pounds_v]=0$ (see, e.g., appendix section A.6. of 
\cite{Besse:2017aa}) the 2D vorticity dynamics may be expressed as a scalar transport equation given by
\begin{equation}\label{eq:2Dvort-trans}
\rmd \omega+ \pounds_{dX_t} \omega= 0 ,
\end{equation}
where we emphasize  because $\omega\in \Omega^0$, in a local chart $(U, \phi)$ with coordinates $x=(x^1,\ldots x^d)$ and $v\in \mathfrak{X}|_{U}$, we have
$
\pounds_{v} \omega_t = v^i \partial_{x^i} \omega_t.
$
Thus, we have the push-forward relation $\omega_t=(\phi_t)_*\omega_0$, where  $\{\phi_t\}$ is the stochastic flow associated with $dX_t$.  Moreover, for all $f \in \Lambda^0$, 
$$
\rmd f(\omega) + \pounds_{dX_t} f(\omega) = 0\,,
$$
and  $f(\omega_t)=(\phi_t)_{*}f(\omega_0)$.
In addition, upon recalling that $\operatorname{div}_g(dX_t)=0$, we find
\begin{equation}\label{eq:cons2Dvort}
\int_{M} f(\omega_t) dA=\int_{M}(\phi_t)_*f(\omega_0) dA = \int_{\phi_t(M)} (\phi_t)_*f(\omega_0) dA= \int_{M}f(\omega_0)\phi_t^* dA=\int_{M}f(\omega_0)dA,
\end{equation}
where, in this section, we $dA$ replaces $dV$ to denote area. 
In particular, choosing $\phi(x)=x^p$ we  find  that all of the $L^p$-norms of the solution are conserved.

We now want to investigate the  fluctuations of the vorticity: 
\begin{equation}\label{eq:flucvort2D1}
\int_{M}  \E{(\omega'_t)^2} dA = \int_{M}  \E{\omega_t^2} dA  - \int_{M} \left(\E{\omega_t}\right)^2 dA.
\end{equation}
We begin by computing the first term $ \int_M \E{\omega_t^2} dV$.  Upon taking $f(x)=x^2$ in \eqref{eq:cons2Dvort}, we find
\begin{equation}\label{eq:expsqvort}
\int_{M}  \omega_t^2 dA = \int_{M}  \omega_0^2 dA \quad \Rightarrow \quad \int _M\E{\omega_t^2} dA = \int_M \E{\omega_0^2} dA.
\end{equation}

Taking the expectation of \eqref{eq:2Dvort-trans}  yields 
\begin{equation}\label{eq:expvort-Ito}
\partial_t\E{ \omega} + \pounds_{\E{u}} \E{\omega}  = \frac{1}{2} \sum_k \pounds_{\xi^{(k)}}\pounds_{\xi^{(k)}} \E{\omega}.
\end{equation}
A  computation  similar to \eqref{eq:diamondcalc} shows that for all $v\in \mathfrak{X}$ with $\operatorname{div}_g v=0$ and $f,g\in \Omega^0$, 
$$
(\pounds_{v} f, g)_{L^2(\Omega^0)}= -( f, \pounds_{v} g)_{L^2(\Omega^0)}
$$
and hence $(\pounds_{v} f, f)_{L^2(\Omega^0)}=0$.  Thus, using the divergence-free property of $dX_t$ yields 
\begin{equation}\label{eq:sqexpvort}
\int_{M}  (\E{\omega_t})^2 dA= \int_{M}  (\E{\omega_0})^2 dA -\sum_k\int_M (\pounds_{\xi^{(k)}}\E{\omega_t})^2 dA,
\end{equation}
which implies that the magnitude  $|\E{\omega}|$ of the expected vorticity will decay to zero in the absence of forcing,  provided that the vector fields $\{{\xi}^{(k)}\}_{k\in \mathbb{N}}$ span $\mathfrak{X}$. 
Therefore, upon substituting  \eqref{eq:expsqvort} and \eqref{eq:sqexpvort} into  \eqref{eq:flucvort2D1}, we find that  fluctuations $\omega'=\omega-\E{\omega}$ satisfy
$$
\int_{M} \E{ (\omega_t')^2}dA=\int_{M}\E{ (\omega_0')^2}dA+\sum_k\int_{M} (\pounds_{\xi^{(k)}} \E{\omega_t})^2 dA,
$$
or equivalently
\begin{equation}\label{eq:vortvar}
\frac{\rmd}{\rmd t} \int_{M} \E{ (\omega_t')^2}dA=\sum_k\int_{M} (\pounds_{\xi^{(k)}} \E{\omega_t})^2 dA.
\end{equation}

This means that the conserved total enstrophy in \eqref{eq:expsqvort} transforms from the mean into the fluctuations for 2D LA SALT vorticity dynamics.  The same phenomenon occurs for the magnitude of the body angular momentum in the finite-dimensional example of rigid-body dynamics, see Section \ref{sec:rigidbody}). 
On the other hand, the total enstrophy itself is preserved because it is a Casimir for the Lie--Poisson structure of the 2D LA SALT Euler equation, given by
\begin{equation}\label{eq:LAvortbrkt}
\rmd C(\omega) = -\int_{M} \omega \,J \bigg( \frac{\delta C}{\delta \omega},
\E{\frac{\delta H}{\delta \omega}}\,  \rmd t + \sum_k \xi^{(k)}\circ \rmd W^{(k)}_t\bigg)dA=0,
\,\end{equation}
with Jacobian operator defined by $J(f,h)dA=\mathbf{d}f\wedge \mathbf{d} h$ just as it is for the deterministic  Euler system. 

\begin{remark}[Synopsis]
Total enstrophy is the spatial integral of the square of vorticity. It is preserved as a result of being a Casimir of the LA SALT system for the Lie--Poisson bracket expressed in \eqref{eq:LAvortbrkt}, so its expectation is preserved by LA SALT. However, consider the enstrophy of the expected vorticity, i.e., the spatial integral of the square of the expectation of vorticity in \eqref{eq:sqexpvort}. According to the results of this paragraph,  the enstrophy of the expected vorticity decays exponentially in time, while  the vorticity variance in \eqref{eq:vortvar} increases exponentially in time.
Consequently, one would expect the probability distribution of enstrophy on a level set of enstrophy in the space of Casimir functionals tends to become more diffuse. This means that as time proceeds the probability density for the enstrophy will tend to a constant on the level set of the enstrophy corresponding to the initial condition. Thus, transport noise in 2D LA SALT vorticity dynamics causes a growth of uncertainty and a loss of  information as the probability distribution for enstrophy delocalises and tends toward the analog for enstrophy of the microcanonical distribution in statistical physics. 

In other words, one may regard the expected vorticity equations for 2D LA SALT in \eqref{eq:expvort-Ito} as a dissipative system embedded into a larger conservative system \eqref{eq:2Dvort-trans}. From this viewpoint, the interaction dynamics of the two components of the full LA SALT system dissipates the enstrophy of the mean vorticity by converting it into fluctuations, while preserving the mean total enstrophy. This dynamics results because the total (mean plus fluctuation) vorticity field is being linearly transported along the mean velocity in \eqref{eq:2Dvort-trans}, while the mean vorticity field  is decaying in 2D dissipative motion \eqref{eq:expvort-Ito}. This is the nature of stochastic coadjoint motion for the LA SALT Euler fluid vorticity equation in 2D, expressed in \eqref{eq:LAvortbrkt}. Namely, the Casimirs are preserved by the full LA SALT dynamics, while the equations for the expected dynamics contain dissipative terms. 
Dissipation in the dynamics of the expected enstrophy is accompanied by loss of information (or increasing uncertainty) as the variance (expected fluctuation enstrophy) grows and the probability distribution of the enstrophy delocalises and tends toward a constant invariant measure. 
\end{remark}

\subsection{Helicity preservation in 3D} \label{sec:Helicity}
The LA SALT Euler fluid motion \eqref{eq:LASALT-Euler} for $u^{\flat}\in \Omega^1$ and  vorticity equation \eqref{eq:LASALT-Euler-Vort}  for $\omega = \mathbf{d}u^{\flat}\in \Omega^2$, together read
\begin{align}\label{LASALT-eqn-press1}
({\rm d}  +    \pounds_{\rmd X_t})u^{\flat}= -\,\mathbf{d}\pi \quad\hbox{and}\quad
({\rm d}  +    \pounds_{\rmd X_t}) \omega= 0\,.
\end{align}
Using that the Lie-derivative satisfies the Leibniz rule with respect to the wedge product, that the exterior derivative $\mathbf{d}$ is an antiderivation, that $\mathbf{d}^2=0$,  and the Stratonovich product rule, we obtain 
\begin{align}\label{LASALT-eqn-press2}
({\rm d}  +    \pounds_{\rmd X_t})(u^{\flat}\wedge \omega )
= -\,\mathbf{d}p\wedge \omega  =  - \mathbf{d}(p \wedge  \omega) 
\,.
\end{align}
Then  using the Kunita-It\^o-Wentzell formula (\cite{Leon:2019aa})  and that $\partial M=\emptyset$, we find
\begin{align*}
\rmd \int_{M} (u^{\flat}\wedge \omega )&=\rmd\int_{\phi_t(M)}(u^{\flat}\wedge \omega )=\int_{M}({\rm d}  +    \pounds_{\rmd X_t})(u^{\flat}\wedge \omega )\\
&=-\int_{M}\mathbf{d}(p \wedge  \omega) =-\int_{\partial M}(p \wedge  \omega)=0,
\end{align*}
where $\{\phi_t\}$ is the stochastic flow corresponding to $dX_t$. 
This integral quantity is known as the \emph{helicity}. Its topological significance as the linkage number for lines of vorticity in a volume preserving fluid flow is discussed by Arnold in \cite{Arnold:1974aa, Arnold:1999aa}. The preservation of helicity for SALT and LA SALT dynamics emphasizes once again the central role played by the Kelvin circulation integral in fluid dynamics.

\subsection{Well-posedness and regularization by non-locality in probability space}\label{Prob-sec}
In this section, we take $M$ to be the flat torus $\mathbb{T}^d=\mathbb{R}^d/\mathbb{Z}^d$.  It is possible to generalize all these results to hold on compact smooth Riemannian manifolds $(M,g)$ without boundary.  For the purpose of introducing notation, let $E$ denote an arbitrary Banach space. Let $L_{\omega}^2E$ denote the Lebesgue Bochner space $L^2(\Omega,\mathcal{F},\mathbb{P};E)$.  For a given time $T>0$ and $p\in [1,\infty]$,  let $L^p_TE$ denote the Lebesgue Bochner space $L^p([0,T],\mathcal{B}([0,T]),\Omega; E)$, where $\Omega$ is the Lebesgue measure on $\mathbb{R}$. For given $d\in \{2,3\}$ and $m\in \mathbb{N}$,  let $H^m_x$ denote the Sobolev space of $m$-times weakly differentiable functions from $\mathbb{T}^d$ to $\mathbb{R}^d$ that have square integrable derivatives up to and including order $m$. We  let $H_{\sigma}^m\subset H^m_x$ denote the subspace of weakly divergence-free functions and $\mathbf{P} \in \mathcal{L}(H^m_x;H_{\sigma}^m)$ denote the bounded linear projection of $H^m_x$ onto $H_{\sigma}^m$ (i.e., the Leray or solenoidal projection). For a given time $T>0$, let $C_TE$ denote the space of strongly continuous functions from $[0,T]$ to $E$. For a given  $m\in \mathbb{N}$, let $C^m_x E$ denote the space of  $m$-times continuously differentiable functions from $\mathbb{T}^d$ to $E$. Let $C^m_{\sigma}E^d\subset C^m_x E^d$ denote subspace of divergence-free  functions.  We let $C^{\infty}_xE=\cap_m C^m_xE$ and similarly we define $C^{\infty}_{\sigma}E^d$. Let $\ell_2^d$ denote the space of square-summable sequences in $\mathbb{R}^d$.  All of the above  spaces introduced above are Banach spaces $V$, and we denote the corresponding norms by $|\cdot|_{V}$. Moreover, we let $(\cdot,\cdot)$ denote the $L^2(\mathbb{T}^d;\mathbb{R}^d)$-inner product and $(\cdot,\cdot)_{H^m_x}=((I-\Delta)^{\frac{m}{2}}\cdot, (I-\Delta)^{\frac{m}{2}}\cdot)$  denote the inner product on $H^m_x$. 

Fix a terminal time $T>0$ and $d\in \{2,3\}$. For a given divergence-free forcing term $f: [0,T]\times \mathbb{T}^d\rightarrow \mathbb{R}^d$, divergence-free initial condition $u_0: \Omega\times \mathbb{T}^d\rightarrow \mathbb{R}^d$, and family of divergence-free vector fields $\{\xi^{(k)}\}_{k\in \mathbb{N}}: \mathbb{T}^d\rightarrow \ell_2^d$,  we consider the equation for an  $\mathbb{F}$-adapted vector field $u:\Omega\times [0,T]\times \mathbb{T}^d\rightarrow \mathbb{R}^d$ and  scalar field. pressure $p: \Omega\times [0,T]\times \mathbb{T}^d\rightarrow \mathbb{R}^d$ given by
\begin{equation}\label{eq:LASALT_Strat}
\begin{cases}
\rmd u_t  +\pounds_{\E{u_t}}^T u_t  \rmd t + \sum_{k}   \pounds_{\xi^{(k)}}^T u_t  \circ \rmd W_t^{(k)}  
= (- \nabla \pi_t + f_t) \rmd t, & \\ 
\operatorname{div}\E{u_t}=0, &  \\
u_t|_{t=0} = u_0,& 
\end{cases}
\end{equation}
where $ \pi_t =  \E{p}$, $\pounds_v^Tu=\operatorname{ad}_v^{\dagger}u=(\pounds_{v}u^{\flat})^{\sharp}= (\operatorname{ad}^*_{v}u^{\flat})^\sharp$ is defined by
\begin{align}\label{e2}
\begin{split}
\pounds^T_{v} u_t  &\coloneqq   v\cdot \nabla u_t + (\nabla v)^T \cdot u_t,
\end{split}
\end{align}
or, more explicitly, as $(\pounds^T_{v} u_t)^i  \coloneqq  v^j\partial_j u^i + (\partial_i v^j)u^j$.   We  interpret \eqref{eq:LASALT_Strat} in the It\^o-formulation:
\begin{align}\label{eq:LASALT_Ito}
\rmd u_t +  \pounds_{\E{u_t}}^T u_t  \rmd t + \sum_{k}  \pounds_{\xi^{(k)}}^T u_t   \rmd W_t^{(k)} &=  \left(\frac{1}{2}  \sum_k \pounds_{\xi^{(k)}}^T( \pounds_{\xi^{(k)}}^T u_t)-\nabla \pi_t  + f_t\right) \rmd t.
\end{align}
A straightforward computation shows that for $\xi,v: \mathbb{T}^d\rightarrow \mathbb{R}^d$,
\begin{align}
\pounds_{\xi}^T(  \pounds_{\xi}^T v) 
&= (\xi\cdot \nabla)\xi\cdot \nabla  v+ (\xi \otimes \xi): (\nabla \otimes \nabla) v+2 \nabla  \xi\cdot   (\xi\cdot \nabla)v+ \nabla ((\xi\cdot \nabla) \xi) \cdot v \nonumber \\
&= \partial_i (a^{ij} \partial_j u^{\alpha}) + b^{i\alpha j }\partial_i u^{j}+ c^{\alpha \beta} u^{\beta},\label{lieComps}
\end{align}
where 
\begin{equation*}
a^{ij} :=   \xi^i\xi^j, \quad b^{i \alpha j }=2\xi^i\partial_\alpha \xi^{j} ,\quad 
c^{\alpha \beta}:=(\partial_{\alpha}\xi^i )\partial_i\xi^{\beta}+\xi^i \partial_{i\alpha }\xi^{\beta},\quad i,j,\alpha,\beta \in \{1,\ldots, d\},
\end{equation*}
and repeated-indices are summed-over.

Taking the expectation of \eqref{eq:LASALT_Ito} yields a closed equation for the deterministic  $v_t= \E{u_t}$ given by 
\begin{align}\label{eq:LLNS}
\partial_t v + {\bf P} \pounds_{v}^T v    &=  {\bf P}   \frac{1}{2}  \sum_k \pounds_{\xi^{(k)}}^T\big( \pounds_{\xi^{(k)}}^T v\big)  + {\bf P}f_t\,.
\end{align}
Equation \eqref{eq:LLNS}  for $\E{u_t}$ generalizes the classical $d$--dimensional Navier-Stokes equations which appear as a special case when $\xi^{(k)}:=\sqrt{2\nu} e_k$, $k=1,2,3,\dots, d$ and  $\xi^{(k)}:=0$ otherwise.  We term these equations \eqref{eq:LLNS} the \emph{Lie-Laplacian Navier-Stokes equations (LL NS)}. 

\vspace{1mm}\noindent\textbf{Assumption 1} There exist a $\kappa>0$ such that  
\be \label{nondegen}
\kappa  |y|^2 \le \frac{1}{2} \sum_k  y^i \xi_i^{(k)}(x)  \xi_j^{(k)}(x)  y^j, \qquad \forall \ x,y\in \mathbb{T}^d.
\ee

\noindent\textbf{Assumption 2}(m) For given $m\in \mathbb{N}$, $u_0\in H^m_\sigma$,   $\xi \in C_\sigma^{m+2}\ell_2^d$, and $f\in L^2_T H_{\sigma}^{m-1}$.

\begin{definition}[Solution of LA SALT Euler]
We say that $u$ is a solution of \eqref{eq:LASALT_Strat} on the interval $[0,T^*]$ if $u$ is a  weakly continuous $H^1_x$-valued $\mathbb{F}$-adapted process  such that $u \in L^2_{\omega} L^{2}_{T^*}H^1_{x}$ and $\E{u}\in L^{2}_{T^*}H^2_{x}\cap L^{\infty}_{T^*}H^1_x$  and for all $\phi \in C^{\infty}_x\mathbb{R}^d$,  $\mathbb{P}$-a.s.\ for all $t\in [0,T^*]$,
\begin{align*}
(u_t,\phi ) &= (u_0,\phi ) + \int^t_0\left[-\sum_k( \pounds^T_{\xi^{(k)}} u_s, \pounds_{\xi^{(k)}}\phi )  + (-\pounds^T_{\mathbb{E} u_s} u_s+f_s)- ( \pi_s,\nabla\cdot \phi) \right] \rmd s \\
&\qquad  - \sum_k\int^t_0(\pounds^T_{\xi^{(k)}} u_s,\phi )\,\rmd  W_s^{(k)},
\end{align*}
where
\begin{equation}\label{def:pressureLASALTEuler}
\pi =(-\Delta)^{-1}\operatorname{div}\left(\operatorname{div}(\E{u}\otimes  \E{u})  -(\pounds_{\xi^{(k)}}^T (\pounds_{\xi^{(k)}}^T \E{u}))\right) +\frac{1}{2} |\E{u}|^2.
\end{equation}
\end{definition}

We remark that the assumption of $u \in L^2_{\omega} L^{2}_{T^*}H^1_{x}$ together with standard embeddings, Jensen's inequality and the convexity of norms give that $\E{u} \in  L^{2}_{T^*}L^4_{x}$ in $d=2,3$.  Therefore $\pi\in L^{1}_{T^*}L^2_{x}$ by elliptic regularity (see proof) and all above terms are well defined.
Our analytical results are as follows. 

\begin{thm}[Well-posedness of LA SALT Euler]\label{thm:LASALTWell} Let $d\in \{2,3\}$ and   Assumption 1 and 2 (m) hold for  $m> \frac{d}{2} + 2$. Then there exists a time $T^*=T^*(m,\kappa^{-1}, |u_0|_{H^m_x}, |f|_{L^2_TH^{m-1}_x}, |\xi|_{C^{m+2}_x\ell_2^d})$ and a unique  solution $u$  of LA SALT on $[0,T^*]$ satisfying  $u\in L^2_{\omega}L^{\infty}_{T^*} H^{n-1}_x $ for $ n := \lfloor m-d/2 \rfloor$. Furthermore,  the solution $u$ is  weakly continuous in time in  $H^{n-1}_x$ and strongly continuous in $H^{n-2}_x$. If $d=2$, then $T^*=\infty$.  Moreover, if $d=3$, there is a  positive number $\kappa^*=\kappa^*(m, |u_0|_{H^m_x}, |f|_{L^2_TH^{m-1}_x}, |\xi|_{C_\sigma^{m+2}\ell_2^d})$ such that for all $\kappa>\kappa^*$,  $T^*=\infty$.
\end{thm}

The proof of Theorem \ref{thm:LASALTWell} proceeds as follows. We first solve \eqref{eq:LLNS} for $v=\E{u}$, and then  solve  the linear equation  \eqref{eq:LASALT_Ito} for $u$. Accordingly, we need a solution theory for the deterministic LL-NS, which we state below.
Note that, owing to \eqref{lieComps} and  \eqref{nondegen},  the Lie-Laplacian   $\frac{1}{2}  \sum_k \pounds_{\xi^{(k)}}^T\big( \pounds_{\xi^{(k)}}^T \cdot\big)$ is the sum of a  divergence-form second-order uniformly elliptic operator $\partial_i (a^{ij} \partial_j \cdot)$ and a non-diagonal and non-symmetric first-order linear differential operator $b^{i}\partial_i + c$. Thus, from the standpoint of well-posedness and solution properties, the arguments to treat LL NS are completely analogous to those used to study the usual Navier-Stokes equation. The only additional ingredient is the solution estimate \eqref{ineq:coerciveNS}. For this reason, we only give a sketch of the proof of Theorem \ref{LLNSthm} and refer reader to the sources \cite{Robinson:2016aa} and \cite{Boyer:2013aa} for more details. 

\begin{thm}[Well-posedness of LL NS]\label{LLNSthm}
Let $d\in \{2,3\}$ and   Assumption 1 and 2 (m) hold for  $m\ge 1$.   Then there exists a time $T^*=T^*(m,\kappa^{-1}, |u_0|_{H^m_x}, |f|_{L^2_TH^{m-1}_x}, |\xi|_{C^{m+2}_x\ell_2^d})$ and a unique strong solution of \eqref{eq:LLNS} on $[0,T^*]$ satisfying $v\in C_{T^*}H^{m}_{\sigma} \cap L^2_{T^*}H^{m+1}_{\sigma}$.   If $d=2$, then $T^*=\infty$.  Moreover, if $d=3$, there is a positive number $\kappa^*=\kappa^*(m, |u_0|_{H^m_x}, |f|_{L^2_TH^{m-1}_x}, |\xi|_{C^{m+2}_x\ell_2^d})$ such that for all $\kappa>\kappa^*$,  $T^*=\infty$.
\end{thm}

\begin{remark}[Loss of regularity in Theorem \ref{thm:LASALTWell}] The loss of regularity the solution  $u$ (i.e., $u$ is not necessarily in $H^m$) is an artifact of our method of proof. We apply a Sobolev solution theory for the linear stochastic transport theorem in which the coefficients in the drift of the SPDE are assumed to be at least $C^1$. We recall that $\pounds^T_{\E{u}} u = \E{u}\cdot \nabla u + (\nabla \E{u})^T\cdot u$, and hence we need at least $\E{u}\in C^2_x\mathbb{R}^d$.  It is plausible that one might be able to avoid the regularity loss by establishing  a regularization by noise result for a linear stochastic transport system or by developing some other non-linear solution scheme. 
\end{remark}

\begin{remark}[Large viscosity and global strong solutions]
It is well-known that a  global strong solution of the Navier-Stokes equation exists for sufficiently small initial data and forcing, or equivalently, sufficiently large viscosity. The lower bound on the viscosity coefficient can be made explicit  in terms of the initial data and forcing, and we refer the reader to \cite{Robinson:2016aa} or Chapter 5 of \cite{Boyer:2013aa} for more details. As explained above, the double Lie-Laplacian is a uniformly elliptic operator with ellipticity constant $\kappa$ (see Assumption 1). Following the method of proof in the case of standard Laplacian dissipation,  one can obtain a lower bound $\kappa^*$ for Lie-Laplacian dissipation depending on the initial data, forcing, and bounds on the coefficients $\xi$.  
\end{remark}

\begin{remark}[Continuity of the solution maps]
Standard techniques can be used to show that the solution map is continuous in initial data, forcing, and noise correlates both for the Lie-Laplacian Navier-Stokes equation
\begin{align*}
S_{LL \ NS}:  H_{\sigma}^{m}\times L^2_{T} H_{\sigma}^{m-1} \times  C_\sigma^{m+2}\ell_2^d &\rightarrow  v \in C_{T^*}H^{m}_{\sigma} \cap L^2_{T^*}H^{m+1}_{\sigma}\\
(v_0, \xi, f) &\mapsto v,
\end{align*}
and, consequently, for the LA SALT Euler equations
\begin{align*}
S_{LA \ SALT}:  H_{\sigma}^{m}\times L^2_{T} H_{\sigma}^{m-1} \times  C_\sigma^{m+2}\ell_2^d &\rightarrow  v \in L^2_{\omega}C_{T^*}H^{n-1}_{x} \cap  L^2_{\omega}L^{\infty}_{T^*} H^{n}_x\\
(u_0, \xi, f) &\mapsto u.
\end{align*}
\end{remark}

\begin{remark}
If $\xi\in C_\sigma^{\infty}\ell_2^d$, $f\in C^{\infty}_T C^{\infty}_{\sigma}$, then $v\in C^\infty(]0, T^*]\times \mathbb{T}^d;\mathbb{R}^d)$ for any $\ve>0$. See  \cite{Robinson:2016aa}, Thm 7.5.
\end{remark}

We now sketch the proofs of these two results.

\begin{proof}[Sketch of Proof of Theorem  \ref{LLNSthm}]

It follows, for example, from  Lemmas 3.6 and 3.7, and  the argument on page 3779 of \cite{Leahy:2015aa}  (see, also, Lemma 5.1 in \cite{Gerencser:2015aa}) that there exist a constant $C=C(d,m,\kappa^{-1},| \xi|_{C^{m+2}_x\ell_2^d})$ such that for all $u\in H^m_x$,
\begin{equation}\label{ineq:coerciveNS}
\frac{1}{2}\sum_k ((\pounds_{\xi^{(k)}}^T)^2 u, u)_{H^m_x}  \le -\kappa' |\nabla u|_{H^m_x}^2 + C|u|_{H^m_x}^2,
\end{equation}
for all $\kappa'<\kappa $.
The proof then follows from a simple modification of standard arguments, see e.g.,  Chapters 6 and 7 of \cite{Robinson:2016aa} or Chapter 5 of \cite{Boyer:2013aa}. Although we do not state them, the dependence of the local existence time and requisite large viscosity for global existence can all be made explicit in terms of $T$ and the size of data and forcing.
\end{proof}

\begin{proof}[Sketch of Proof of Theorem  \ref{thm:LASALTWell}]
Owing to Theorem 3.3 in \cite{Leahy:2015aa} (see, also, Theorem 3.1 in \cite{Gerencser:2015aa}), if $u_0 \in H^n_x$, $\E{u}\in L^{\infty}_{T^*}C^{n+1}_x$ , $\xi  \in C^{n+2}_x\ell_2^d$,  and $f,\nabla \pi\in L^2_{T^*} H^{n}_x$, then there exists a    solution $u$ of LA SALT on the interval $[0,T]^*$ such that $u\in L^2_{\omega}L^2_{T^*}H^n_x$. 
Moreover, $u$ is weakly continuous in $H^n_x$ and strongly continuous in $H^{n-1}_x$.  These results are based on the following a priori estimate: there is a constant $C=C(d,m,|\E{u}|_{L_{T^*}^{\infty}C^{n+1}_x},|\xi|_{C_x^{n+1}\ell_2^d})$ such that for all $u\in H^n$, 
\begin{equation}\label{ineq:a priori linear}
(\pounds_{\E{u}}^Tu,u)_{H^n_x}+\frac{1}{2}\sum_k\left((\pounds_{\xi^{(k)}}^T)^2 u, u\right)_{H^n_x} + \frac{1}{2}\sum_k|\pounds_{\xi^{(k)}}^Tu|_{H^n_x}^2 \le C |u|_{H^n_x}^2.
\end{equation}

By virtue of the Sobolev embedding theorem and Theorem  \ref{LLNSthm},  if  $u_0\in H^{m}_{\sigma}$, $f\in L^2_TH^{m-1}_{\sigma}$, and $\xi \in C^{m+2}_{\sigma}\ell_2^d$  for $m> d/2 +n+1$, then $\E{u}\in L_{T^*}^\infty C^{n+1}_{\sigma}$.

Taking the divergence of both sides of \eqref{eq:LASALT_Ito}, we obtain the  expression \eqref{def:pressureLASALTEuler} for the expected pressure. 
Using standard estimates of elliptic PDE in Sobolev spaces (see, e.g., Thm III 4.1 and 4.2 in \cite{Boyer:2013aa}), we find 
$$
|\pi|_{H^{m-1}_x}  \le \frac{3}{2}|\E{u}\otimes \E{u}|_{H^{m-1}_x} + \left|\operatorname{div}\left((\pounds_{\xi^{(k)}}^T (\pounds_{\xi^{(k)}}^T \E{u}))\right)\right|_{H^{m-3}_x}.
$$
Noting that $m -1> n + \frac{d}{2}$, by the Banach-algebra  property of $H^{m-1}_x$ (see, e.g., Lemma 3.4 in \cite{Majda:2002aa}), there is a constant $C=C(d,m)$ such that 
$
|\E{u}\otimes \E{u}|_{H^{m-1}_x}\le C |\E{u}|_{H^{m-1}_x}^2.$
Moreover, since $\operatorname{div} \E{u}=0$, it follows from Lemma 3.6 and  the argument on page 3778 of \cite{Leahy:2015aa} that  there is a constant $C=C(d,m, |\xi|_{C^m_x\ell_2^d})$ such that 
$
\left|\operatorname{div}\left((\pounds_{\xi^{(k)}}^T)^2\E{u}\right)\right|_{H^{m-3}_x} \le   C|\E{u}|_{H^{m-1}_x}.
$
Thus, we obtain
$$
\int_0^{T^*}|\nabla\pi_t|^2_{H^n_x}dt \le \int_0^{T^*}| \pi_t|^2_{H^{m-1}_x}dt \le C\int_0^{T^*} |\E{u_t}|_{H^{m-1}_x}^2dt<\infty,
$$
which gives   $\nabla \pi  \in L^2_{T^*} H^{n}_x$.

We now turn our attention to uniqueness. If $u_1,u_2$ are solutions of LA SALT, then $\E{u_1}$ and $\E{u_2}$ are  strong solutions of LL-NS.  Thus, $\E{u_1}= \E{u_2}$ since strong solutions are unique by Theorem \ref{LLNSthm}.  It then follows from the uniqueness of linear stochastic transport equations that $u_1=u_2$. 
\end{proof}

\begin{remark}
In fact, estimates of the form \eqref{ineq:a priori linear} hold for more general tensor fields.  Lemma 3.7 along with the argument on page 3779 of \cite{Leahy:2015aa}  (see, also, Lemma 5.1 in \cite{Gerencser:2015aa}) directly imply that for all vector fields $v\in \mathfrak{X}$, there is constant $C=C(d,m,|v|_{C^{m+2}})$ such that for all $\tau \in \mathcal{T}_s^r$, we have
\begin{equation}\label{ineq:tensor}
\left(\pounds_{v}^2 \tau, \tau\right)_{H^m( \mathcal{T}_s^r)} + |\pounds_{v} \tau|_{H^m( \mathcal{T}_s^r)}^2 \le C |\tau|_{H^m( \mathcal{T}_s^r)}^2
\end{equation}
and for all $\alpha \in \Omega^k$, 
\begin{equation}\label{ineq:kform}
\left(\pounds_{v}^2 \alpha, \alpha\right)_{H^m( \Omega^k)} + |\pounds_{v} \tau|_{H^m(\Omega^k)}^2 \le C |\tau|_{H^m(\Omega^k)}^2.
\end{equation}
Indeed,  estimates of this form are derived in \cite{Leahy:2015aa} and \cite{Gerencser:2015aa}  for a  first-order differential operator acting on functions $\phi : \mathbb{T}^d\rightarrow \mathbb{R}^{d'}$  of the form 
$$(\mathcal{L}\phi)^{\alpha}  = v^j \partial_j \phi^{\alpha}+ \Omega^{\alpha j} \phi^{j}, $$ where $\xi\in C^{m+1}(\mathbb{T}^d;\mathbb{R}^d)$ and $ \Omega \in C^{m+1}(\mathbb{T}^d;\mathbb{R}^{d'\times d'})$. That is, there is constant $C=C(d,m,|v|_{C^{m+1}})$ such that \eqref{ineq:tensor}  holds with $\pounds_v$ replaced with $\mathcal{L}$ and $\tau$ or $\alpha$ replaced with $\phi$. 
One can see that the highest-order part (first-order) of $\mathcal{L}$ acts diagonally, which is why the Lie-derivative is a particular case. We remark also that estimates of this type were first derived for scalar functions in Lemma 2.1 of \cite{Krylov:1986aa} (see, also, Lemma 4.3 in \cite{Rozovsky:2018aa}). In fact, estimates given in Lemma 5.1 of \cite{Gerencser:2015aa} generalize such estimates to $L^p$-Sobolev spaces. 
\end{remark}

\begin{remark}[Method of characteristics solution and representation]\label{moc}
Let $\alpha>0$. If $b\in L_{T^*}^{\infty} C^{2+\alpha}_x$ and $\xi\in C^{3+\alpha}_x\ell_2^d$, then there exists a stochastic flow of $C^{2+\alpha'}_x$-diffeomorphisms $\phi=\{\phi_{s,t}\}$  for $\alpha'<\alpha$ satisfying 
$$
d\phi_{s,t}(x) =b(\phi_{s,t}(x)) \rmd t+\sum_k \xi^{(k)}(\phi_{s,t}(x)) \circ \rmd W_t^{(k)}, \quad \phi_{s,s}(x)=x\in \mathbb{T}^d.
$$  
We take $b=\E{u}$, which is sufficiently regular provided Assumption 1 and 2 (m) hold for $m>\frac{d}{2}+2+\alpha$ by Theorem \ref{LLNSthm}.
We denote the  spatial inverse of the flow (i.e, the back-to-labels map) by $A_t=\phi^{-1}_{0,t}$. The inverse  can be shown directly to satisfy (see, e.g., Theorem 2.4 in \cite{Leahy:2015ab})
$$
\rmd A_t(x) = \E{u_t}(x)\cdot \nabla A_t(x) \rmd t + \sum_k \xi^{(k)}(x)\cdot \nabla A_t(x)\circ \rmd W_t^{(k)}, \quad A_0(x)=x.
$$
It is easy to verify that  $J_t^{-1}=(\nabla \phi_{0,t})^{-1}$ satisfies 
$$
\rmd J_t^{-1}(x) = -J_t^{-1}(x)\nabla \E{u_t}(\phi_t(x)) \rmd t -J_t^{-1}(x) \sum_k  \nabla \xi^{(k)}(\phi_t(x))\circ 
\rmd W_t^{(k)}, \quad J_0^{-1}(x)=I.
$$
Noting that $J_t^{-1}(A_t)=\nabla A_t$, we find (see, e.g., Theorem 2.2 in \cite{Leahy:2016aa}))
$$
u_t(x)= (\nabla A_t(x))^T\left[u_0(A_t(x))+\Psi_t(A_t(x))\right], \quad  
\Psi_t(x)=\int_0^t J_s^{-1}(x)\left(f_s(\phi_s(x)) + {\nabla \pi}_t(\phi_s(x))\right)\rmd s.
$$
In geometric notation, we may write the above as 
\begin{equation}\label{eq:flowrep}
u^{\flat}_t(x)= (\phi_{0,t})_*\left(u_0^{\flat}(x) + \psi^{\flat}(x)\right),
\end{equation}
where $\psi^{\flat}$ is  defined in terms of $f^{\flat}$ and $\E{d p}$, where $dp\in \Omega^1$ is exterior differential of $p$. 
This representation is closely related  to the representation derived for the SALT system without a forcing term that is discussed in Remark 6 of \cite{Drivas:2018aa}. 
In fact, in \cite{Drivas:2018aa}, the authors  derive a stochastic representation for the solution $\E{u}$ of  \eqref{eq:LLNS} akin to the Constantin-Iyer representation \cite{Constantin:2008aa}:
\be\label{Cipre}
\E{u_t}=\E{\mathbf{P}\left[(\nabla A_t(x))^T u_0(A_t(x))\right]}.
\ee
In fact, in those works, $u_t$ is identified as a stochastic Weber velocity and the representation \eqref{Cipre} can directly be obtained from \eqref{eq:flowrep} for the `Weber velocity' by taking expectation.
\end{remark}

\section{Further examples of LA SALT Systems}\label{sec:examples}

This section illustrates the applicability of the LA SALT approach for a wide range of dynamical equations. 
Section \ref{sec:rigidbody} discusses a finite dimensional example at the foundations of geometric mechanics. This is the rigid body in three dimensions. This example is a great source of intuition and intuition for dynamics in geometric mechanics, and we expect it to be foundational in the exploration of applications of SALT and LA SALT, as well. See \cite{Cruzeiro:2018aa} for the general theory of SALT for Euler-Poincar\'e equations in finite dimensions. The corresponding general theory for LA SALT remains to be established in finite dimensions. 

Section \ref{sec:Burgers} treats the simplest LA SALT fluid equation. This is the Burgers equation, a fundamental 1D nonlinear  equation which has been used to develop insight and intuition throughout fluid dynamics. Here we sketch an open problem for the LA SALT version of the Burgers equation. Namely, what is the signature of the noise correlates $\xi^{(k)}$ in  the zero noise limit of LA SALT Burgers equation? 

Section \ref{sec:CHeqn} discusses the LA SALT version of the Camassa-Holm (CH) equation for singular nonlinear shallow water waves which are effectively finite dimensional.  The LA SALT CH equation is the simplest example of the LA SALT formulation for soliton equations, yet it also summons fundamental questions about the regularity of its solutions. 

Finally, to illustrate the breadth of applicability of the LA SALT formulation, we discussion its application for the multi-physics problem of  incompressible, vertically stratified magnetohydrodynamics (MHD) in three dimensions in section \ref{sec:LASALT-MHD}. The MHD example is meant to sketch yet another open problem and to show that the LA SALT formulation can accommodate a wide range of fluid applications. We expect the pursuit of these open problems in the applications of LA SALT to be fruitful and useful for fluid dynamics. 

\subsection{LA SALT rigid body: a finite dimensional example without advection}\label{sec:rigidbody}

In this example, we take $\mathfrak{X}=\mathfrak{so}(3)\cong \mathbb{R}^3$ and $\mathfrak{X}^*=\mathfrak{so}^*(3)\cong\mathbb{R}^3$ with the dot-product pairing (i.e., there is no density component $\mathfrak{X}^*$).  Upon choosing
$\ell(\mathbf{\Omega})=\frac{1}{2}\mathbf{\Omega}^T \cdot I\mathbf{\Omega},$ where ${\rm I}={\rm diag}({\rm I}_1,{\rm I}_2,{\rm I}_3)$ is the moment of inertia in the body frame, we find $\mathbf{\Pi}:=\frac{\delta \ell}{\delta \mathbf{\Omega}}=I \mathbf{\Omega}$. As explained in  \cite{Arnaudon:2018aa}, the SALT formulation of  the stochastic rotation of a rigid body in $\mathbb{R}^3$ is governed  by 
\begin{align}\label{SRB-Strat}
{\rm d} \mb{\Pi} + \Big( \mb{\Omega}{\rm d}t + \sum_k \xi^{(k)} \circ {\rm d}W_t^{(k)} \Big) \times \mb{\Pi}
= 0\,,
\end{align}
where $\mb{\Pi}$ is interpreted as the angular momentum vector and  $\mb{\Omega}={\rm I}^{-1}\mb{\Pi}$ as the angular velocity vector.
If we replace  $\mb{\Omega}$ by the expected angular velocity $\E{ \mb{\Omega}}$,  we obtain the LA SALT body dynamics
\begin{align}\label{SRB-Strat1}
{\rm d} \mb{\Pi} + \E{ \mb{\Omega}}\times \mb{\Pi} \,{\rm d}t 
+ \sum_k \xi^{(k)} \times \mb{\Pi}  \circ {\rm d}W_t^{(k)} = 0
\,.
\end{align}
In either the SALT or LA SALT body dynamics, one has
\be
\frac{\rmd }{\rmd t}| \mb{\Pi} |^2 = 0,
\label{am-sphere}
\ee
which follows from \eqref{SRB-Strat1} and the permutation vector identity $a\cdot b\times c = c \cdot a \times b$.  Therefore, the probability distribution for the SALT and LA SALT rigid body lies on a level set of $|\mb{\Pi}|^2$. Thus, these dynamics represent stochastic coadjoint motion on a level set of the Casimir function of the Lie--Poisson bracket. 

In the It\^o representation, this stochastic motion equation \eqref{SRB-Strat1} becomes 
\begin{align}
{\rm d} \mb{\Pi}+\Big( \E{ \mb{\Omega}} {\rm d}t  + \sum_k \xi^{(k)}  {\rm d}W_t^{(k)} \Big) \times \mb{\Pi}
= \frac12 \sum_k \xi^{(k)}\times \big(\xi^{(k)} \times \mb{\Pi}\big) {\rm d}t
\,.
\label{SRB-Ito-Pi}
\end{align}

Taking the expectation yields
\begin{align}\label{SRB-Ito-Ex1}
\frac{{\rm d}}{{\rm d}t} \E{\mb{\Pi}} + \E{ \mb{\Omega}}  \times \E{\mb{\Pi}} 
= \frac12 \sum_k \xi^{(k)}\times \big(\xi^{(k)} \times \E{\mb{\Pi}} \big). 
\end{align}
Upon again using the permutation vector identity, we find 
\begin{align}\label{SRB-Ito-Ex2}
\frac{{\rm d}}{{\rm d}t}  \big|\E{\mb{\Pi}}\big|^2 =
- \sum_k  \big|\xi^{(k)} \times \E{\mb{\Pi}} \big|^2{\rm d}t
\,.
\end{align}
Thus, the magnitude $|\E{\mb{\Pi}}|$ of the expected body angular momentum vector $\E{\mb{\Pi}}$ will decay to zero in the absence of forcing. This means that $\E{\mb{\Pi}}$ itself and $\E{\mb{\Omega}}$ will also decay to zero, provided that $\{\xi^{(k)}\}_{k\in \mathbb{N}}$ span $\mathbb{R}^3$.
To calculate the fluctuation dynamics of $\mb{\Pi}' :=\mb{\Pi}- \E{\mb{\Pi}}$, we subtract equation \eqref{SRB-Ito-Ex1} from equation \eqref{SRB-Strat1} to find
\begin{align}\label{SRB-PiFluct}
{\rm d} \mb{\Pi}' + \E{ \mb{\Omega}}\times \mb{\Pi}' \,{\rm d}t 
+ \sum_k \xi^{(k)} \times \mb{\Pi} \rmd W_t^{(k)} 
= \,\frac12 \sum_k \xi^{(k)}\times \big(\xi^{(k)} \times \mb{\Pi}' \big){\rm d}t
\,.
\end{align}
By the It\^o product rule we have
\be
\frac12{\rm d} |\mb{\Pi}'|^2 = \mb{\Pi}' \cdot  \rmd \mb{\Pi}' + \frac{1}{2} \rmd \langle \mb{\Pi}', \mb{\Pi}' \rangle_t=  \mb{\Pi}' \cdot  \rmd \mb{\Pi}' +\frac12\sum_k  \big|\xi^{(k)} \times \mb{\Pi}  \big|^2 \rmd t.
\ee
Thus, upon taking the dot product of $ \mb{\Pi}'$ with \eqref{SRB-PiFluct}, applying the It\^o cross-variance formula implies
\begin{align*}
\frac12{\rm d} |\mb{\Pi}'|^2 + \mb{\Pi}' \cdot \sum_k \xi^{(k)} \times \mb{\Pi} \,{\rm d}W_t^{(k)} 
&=
\frac12\sum_k \Big(  \mb{\Pi}' \cdot  \xi^{(k)} \times \big(\xi^{(k)} \times  \mb{\Pi}' \big) 
+ \big|\xi^{(k)} \times \mb{\Pi}  \big|^2 \Big)
\\&= \frac12\sum_k \Big( - \big|\xi^{(k)} \times \mb{\Pi}'  \big|^2 + \big|\xi^{(k)} \times \mb{\Pi}  \big|^2 \Big)\\
&= \frac12\sum_k \Big( 2 (\xi^{(k)}\times\E{\mb{\Pi} })\cdot (\xi^{(k)}\times \mb{\Pi}) - |\xi^{(k)}\times \E{\mb{\Pi} }|^2\Big).
\end{align*}
Taking the  expectation then yields
$$
\frac{{\rm d}}{{\rm d}t} \E{ |\mb{\Pi}'|^2} = \sum_k |\xi^{(k)} \times \E{\mb{\Pi} }|^2,
$$
and hence the fluctuations grow as time increases.

In summary, level sets of $|\mb{\Pi}|^2$ are preserved, by \eqref{am-sphere}.
That is, the invariant measure for the motion is supported on the angular momentum sphere. This is the same result as for the SALT rigid body, treated in \cite{Arnaudon:2018aa}.  On the other hand, for LA SALT the variances grow monotonically, so the distribution of angular momentum on the angular-momentum sphere  tends to become more diffuse. Consequently, one expects the probability distribution for the magnitude of the angular momentum of LA SALT rigid body to tend toward a constant on a level set of $|\mb{\Pi}|^2$. This was proved to happen as well for the SALT rigid body in \cite{Arnaudon:2018aa}.

\subsection{The LA SALT Burgers equation}\label{sec:Burgers}
Fix, for simplicity  $M=\mathbb{S}^1$.
Choosing $\ell(u)=\frac{1}{2}\int_{S^1}|u|^2\,dx$, the one dimensional LA SALT Burgers equation reads
\be
\rmd u  +\E{u_t} \partial_x u \,\rmd t+  \sum_{k} {\xi}^{(k)}\partial_x u  \circ \rmd W_t^{(k)}    = 0.
\ee
The SALT Burgers equations (without the expectation on the drift velocity) were studied in \cite{Alonso-Oran:2018aa} and it was shown that  shocks form almost surely.  On the other hand, the solutions of the LA SALT Burgers equation stay regular.  Indeed, the expectation  $v_t= \mathbb{E}[u_t]$ satisfies
\be
\partial_t v  +v \partial_x v     = \sum_{k}  {\xi}^{(k)}\partial_x ({\xi}^{(k)}\partial_x  v)
\,,\ee
which is a viscous Burgers equation.
Thus, if the $\xi^{(k)}$ are sufficiently smooth and non-degenerate \eqref{nondegen}, then the above equation gives rise to a global smooth solution $v_t$. The full field is then recovered by a linear transport equation, as in the LA SALT Euler case. Thus, Burgers equation provides a clear example of regularization by `non-locality' in probability space.
Note also from the transport structure one has the representation 
\be\label{burgrep}
u_t(x) = u_0(A_t(x)), \qquad v_t = \mathbb{E}[u_0(A_t(x))]
\,,\ee
where $A_t$ is the back-to-labels map defined in Remark \ref{moc}.
At the level of the mean $v_t$,  the above reasoning generalizes the stochastic method of characteristics (Feynman-Kac formula) for the usual Burgers equation \cite{Constantin:2008aa}.  This representation was used in the work \cite{eyink2015spontaneous} to study the limit of vanishing viscosity for viscous Burgers.  There, it was shown the Lagrangian trajectories in the zero noise/viscosity limit become non-unique backward in time due to stochastic splitting that occurs at shock points.  This phenomenon, known as spontaneous stochasticity, has many implications for understand high-Reynolds number, turbulent physics such as Richardson particle dispersion \cite{bernard1998slow,eyink2013flux,lazarian2015turbulent,lalescu2015inertial}, anomalous dissipation \cite{g00phase,falkovich2001particles,Drivas:2017aa,drivas2017lagrangian2}, and time-irreversibility \cite{eyink2006turbulent, jucha2014time,drivas2019turbulent}.    It would be interesting to study the effect of the $\xi^{(k)}$ functions on the zero-noise limit of LA-SALT Burgers, as well as their signature on the spontaneously stochastic probability measure on generalized trajectories of the entropy solutions.

\subsection{LA SALT Camassa--Holm (CH) equation}\label{sec:CHeqn}
Fix, for simplicity  $M=\mathbb{S}^1$.
Upon choosing $\ell(u)=\frac{1}{2}\int_{\mathbb{S}^1} \left(u^2 + \alpha^2 (\partial_x u)^2 \right)dx$ and defining the Lie derivative of a one-form density $m = m (dx)^2$ by a vector field $\xi$ in 1D as
\begin{align}\label{Exp-mom1}
\pounds_\xi m = (\partial_x m + m \partial_x)\xi \,,
\end{align}
one finds the LA SALT CH equation
\begin{align}\label{om-SPDE2D-brkt-Sum-Strat}
\begin{split}
{\rm d} m &= - (\partial_x m + m\partial_x) \Big( \E{\frac{\delta H}{\delta m }} dt 
+ \sum_{k} \xi^{(k)} \circ {\rm d}W_t^{(k)} \Big) 
\\&=  - \pounds_{K*\E{m}} m \,\rmd t - \sum_{k}  \pounds_{\xi^{(k)}}  m \circ {\rm d}W_t^{(k)}
\,.
\end{split}
\end{align}
where $K(x)=\frac12 \exp (-|x|/\alpha)$ is the Green's function for the 1D Helmholtz operator $1- \alpha^2 \partial_x^2$ and
where the stochastic CH  Hamiltonian  is given by 
\begin{align}\label{stoch-HamCH}
dh(m) = H(m) \rmd t +\sum_{k}  \int_{\mathbb{S}^1} m \xi^{(k)} \circ {\rm d}W_t^{(k)} dx,
\end{align}
with deterministic part 
\begin{align}\label{det-HamCH}
H(m) = \frac12 \int_{\mathbb{S}^1} m \,K*m dx
= \frac12 \int_{\mathbb{S}^1} \int_{\mathbb{S}^1}  m(x) \,K(|x-x'|)m(x') dx' dx
\,.
\end{align}
The expected velocity $\E{u}=\E{\delta H/\delta m} $ is given in terms of expected momentum $\E{m}$ by $\E{u} = K*\E{m}$,
with $m = u - \alpha^2 u_{ xx}$ and $u=K*m$.

As an It\^o stochastic transport equation, \eqref{om-SPDE2D-brkt-Sum-Strat} reads
\begin{align}\label{om-SPDE2D-brkt-Sum-Ito}
{\rm d} m =  - \pounds_{K*\E{m}} m \,\rmd t - \sum_{k}  \pounds_{\xi^{(k)}}  m \, {\rm d}W_t^{(k)}
+ \frac12 \sum_{k}  \pounds_{\xi^{(k)}} \big( \pounds_{\xi^{(k)}} m \big) \rmd t
\,,
\end{align}
and its expectation immediately yields the dissipative equation 
$$
\partial_t\E{m} =  - \pounds_{K*\E{m}} \E{m} 
+ \frac12 \sum_{k}  \pounds_{\xi^{(k)}} \big( \pounds_{\xi^{(k)}} \E{m} \big)
\,.
$$

The subsequent calculations for LA SALT CH would follow the path established in section \ref{sec:LiePoisHam} for deriving the dynamics of the fluctuations \eqref{eq:LASALT-EP-Fluc-Strat} and the spatially integrated variance \eqref{eq:LASALT-EP-var} for the LA SALT CH equation. Rather then follow that path here, though, we shall consider the reduction to a finite dimensional system of SDEs which are nonlocal in probability space, arising from the singular momentum map afforded by the Lie--Poisson structure for the LA SALT equation in \eqref{om-SPDE2D-brkt-Sum-Strat}  \cite{Holm:2005aa}.
\vspace{2mm}

{\bf LA SALT CH Peakons.} The Stratonovich version of the LA SALT CH equation \eqref{om-SPDE2D-brkt-Sum-Strat} admits \emph{singular solutions} for the 1D momentum $m_t$. In particular, these singular solutions can be distributions of momentum on points in the real line.  In previous work for the SALT version of the CH equation, the singular solutions (peakons) were found to form with positive probability \cite{Crisan:2018aa}. The singular peakon solution Ansatz is given by \cite{Camassa:1993aa}:
\begin{align}\label{singMomap}
m_t(x) = \sum_{a=1}^N p_a(t) \delta (x-q_a(t)) 
\quad\hbox{and}\quad 
u_t(x) = \sum_{a=1}^N p_a(t) K (x-q_a(t)) 
\,,
\end{align}
with $K(x)=\frac12 \exp (-|x|/\alpha)$. Substitution of the peakon solution Ansatz into the LA SALT CH equation \eqref{om-SPDE2D-brkt-Sum-Strat} yields the following closed SDEs for the time-dependent parameters $q_a(t)$ and $p_a(t)$, 
\begin{align}\label{Peakon-eqns}
\begin{split}
\rmd q_a &= \E{u_t(x) \Big|_{x=q_a}} dt + \sum_{k} \xi^{(k)}(q_a) \circ {\rm d}W_t^{(k)} 
\,,\\ \\
\rmd p_a &= - p_a \E{ \frac{\partial u_t(x)}{\partial x}  \bigg|_{x=q_a}} dt 
- p_a \sum_{k} \frac{\partial \xi^{(k)}(x)}{\partial x} \bigg|_{x=q_a} 
\hspace{-3mm}\circ {\rm d}W_t^{(k)} 
\,,\end{split}
\end{align}
with $u_t(x)$ given in \eqref{singMomap}.

\begin{remark}
Although peakons have been shown to emerge in the initial value problem for SALT CH with positive probability \cite{Cruzeiro:2018aa}, the issue of whether peakons emerge for the LA SALT CH dynamics in \eqref{om-SPDE2D-brkt-Sum-Strat} from confined initial conditions for velocity $u(x,0)$ remains an open question at this time. However, if the initial condition contains \emph{only peakons}, then it's clear from equation \eqref{Peakon-eqns} that they persist, so long as the solution exists for their dynamics governed by the closed system of SDEs in \eqref{Peakon-eqns} for any finite number of peakons. These interesting, but unfamiliar LA SALT CH peakon SDEs have yet to be studied.
\footnote{See \cite{bendall2019perspectives} for numerical simulations of these equations. These simulations show the emergence of trains of peakons whose amplitudes (speeds) tend to decay in time.}
\end{remark}

\subsection{Incompressible, vertically stratified 3D LA SALT magnetohydrodynamics (MHD)}\label{sec:LASALT-MHD}$\,$\medskip
Fix, for simplicity  $M=\mathbb{T}^3$.
To formulate a comprehensive example, we consider the LA SALT MHD equations for an incompressible stratified medium moving in three dimensions under constant acceleration of gravity, $g$. The corresponding equations for 3D LA SALT MHD are given by, 
\begin{align}\label{mhd-SPDE3D}
\begin{split}
\rmd  u + ({\rm d}X_t \cdot\nabla)  u + u_j \nabla \rmd X_t^j 
&= - \nabla\Big( \E{p} - \frac12 \E{| u|^2}\Big)\,\rmd t  - g\E{b}{\hat{z}}\,\rmd t  
\\&\qquad + gz\nabla \big(b - \E{b}\big)\,\rmd t  + \E{J}\times B\,\rmd t 
\,,\\
\rmd b + {\rm d}X_t \cdot\nabla b = 0 \,, 
&\quad 
\rmd B - {\rm curl} \big({\rm d}X_t \times B \big) 
+ {\rm d}X_t({\rm div\,} B) = 0
\,.
\end{split}
\end{align}
In these equations, the stratification is measured by buoyancy, $b$. The magnetic field $B$ is divergence free, so that ${\rm div} B = 0$. The current density is given in terms of the magnetic field by $J:= {\rm curl}B$. Finally, the  transport velocity is given by the LA SALT Stratonovich stochastic vector field,
\begin{align}\label{mhd-dx3D}
{\rm d}X_t  = \E{ u}(x )\,\rmd t + \sum_{k} \xi^{(k)}(x)\circ \rmd W_t^{(k)} 
\,.
\end{align}
We note that the constraint ${\rm div} B = 0$ is preserved, if it holds initially, which we will assume henceforth.  Also, it's clear that writing the corresponding equations in It\^o form would be straightforward, given the amount of previous description above. 
The physical variables for 3D incompressible MHD are: momentum $\mu\in \Omega^1\otimes \operatorname{Den}$, mass density $D\in \operatorname{Den}$, buoyancy $b\in \Omega^0$ and magnetic flux $B \in \Omega^2$, with components,
\begin{align}
\mu =  \mu\cdot d \mb{x}\otimes d^3x
\,,\quad D=\rho\, d^3x
\,,\quad b = b
\,,\quad \hbox{and}\quad 
B = B\cdot d S
\,.
\end{align}
That is, $V= \operatorname{Den}\otimes \Omega^0\otimes \Omega^2$ and $a= (D,b,B)\in V$.
The Hamiltonian for deterministic 3D incompressible, vertically stratified MHD in terms of the physical variables is
\begin{align}
H(\mu, D,b, B) = \int_{\mathbb{T}^3} \left(\frac{1}{2\rho}  | \mu|^2 +g\rho bz + \frac12 |B|^2  + p(\rho-1)\right)\,d^3x
\,,
\end{align}
whose variational derivatives are given by
\begin{align}
\delta H(\mu, D,b, B) = \int_{\mathbb{T}^3}\left(\frac{ \mu}{\rho}\cdot \delta  \mu
+ \Big(p -  \frac{| \mu|^2}{2\rho^2}  + gbz\Big)\,\delta \rho + g\rho z\,\delta b
+ B\cdot \delta B \right)\,d^3x
\,,
\end{align}
so that one finds $ u= \mu/\rho$. 

The entries in the Hamiltonian operator in equation \eqref{SDP-LASALT-MHD} for 3D incompressible vertically stratified MHD are given in the first column by
\begin{align}
\begin{split}
\pounds_v (\rho\, d^3x) &= {\rm div}(\rho v)\, d^3x
\,,\quad 
\pounds_v (\mu_i dx^i \otimes d^3x) 
= \big( (\partial_j \mu_i + \mu_i \partial_j )v^j \big)dx^i  \otimes d^3x
\\
\pounds_v b &=  v\cdot\nabla b
\,,\quad 
\pounds_v (B \cdot d S) = \big(-{\rm curl\,}( v\times B )
+  v ({\rm div\,} B) \big)\cdot dS
\,.
\end{split}
\label{key-relations}
\end{align}
Then, by the definition of the diamond operator $\diamond: V^*\times V\to \mathfrak{X}^*$ in terms of the Lie derivative in \eqref{diamond-def}, the remaining entries in the first row of the Hamiltonian operator involving diamond $(\diamond)$ are given by 
\begin{align}\label{diamonds}
\frac{\delta H}{\delta D} \diamond D = D \nabla \frac{\delta H}{\delta D}
\,,\quad 
\frac{\delta H}{\delta b} \diamond b = - \frac{\delta H}{\delta b} \nabla b 
\,,\quad 
\frac{\delta H}{\delta  B} \diamond  B 
=   B\times {\rm curl\,}\frac{\delta H}{\delta  B}  - \frac{\delta H}{\delta  B} {\rm div\,} (B) 
\,.
\end{align}
These relations are sufficient to develop the dynamical equations \eqref{mhd-SPDE3D} of the LA SALT 3D incompressible MHD example in Hamiltonian matrix form, as 
\begin{align}\label{SDP-LASALT-MHD}
{\rm d} 
\begin{bmatrix}
\mu \\ \\  D \\ \\  b \\ \\  B
\end{bmatrix}
= -
\begin{bmatrix}
\pounds_{(\,\cdot\,)}\mu & (\,\cdot\,)\diamond D
& (\,\cdot\,)\diamond b  & (\,\cdot\,)\diamond B
\\ \\  
\pounds_{(\,\cdot\,)} D       & 0    & 0    & 0
\\ \\  
\pounds_{(\,\cdot\,)} b           & 0    & 0    & 0
\\ \\  
\pounds_{(\,\cdot\,)} B           & 0    & 0    & 0
\end{bmatrix}
\begin{bmatrix}
\E{ \delta H / \delta\mu } {\rm d}t + \sum_k {\xi}^{(k)} \circ {\rm d}W_t^{(k)} 
\\ \\  
\E{ \delta H/ \delta D} {\rm d}t 
\\ \\  
\E{ \delta H/ \delta b} {\rm d}t 
\\ \\  
\E{ \delta H/ \delta B} {\rm d}t 
\end{bmatrix}
.\end{align}

The Casimirs for 3D vertically stratified incompressible MHD are 
\begin{align}\label{Cas-LASALT-MHD}
C[b,B] = \int_{\mathbb{T}^d}\rho\, \Phi(b, \rho^{-1}B\cdot\nabla b ) \,d^3x\,,
\end{align}
for any differentiable function $\Phi\in \Lambda^0$. 

Equations \eqref{key-relations}--\eqref{SDP-LASALT-MHD} deliver the system \eqref{mhd-SPDE3D} for 3D incompressible, stratified LA SALT MHD in the framework established in section \ref{sec:LiePoisHam} for obtaining the complete dynamics of the expected solutions in equation \eqref{eq:LASALT-EP-Exp}, the fluctuations in equation \eqref{eq:LASALT-EP-Fluc-Strat} and the variances in equation \eqref{eq:LASALT-EP-var}. The semidirect-product Lie--Poisson Hamiltonian operator in equation \eqref{SDP-LASALT-MHD} is identical to the corresponding operator for the deterministic case investigated in \cite{Holm:1998aa}. 

\begin{appendices}

\section{Historical background of the geometric approach to fluid mechanics}\label{Arnold-sec}

In two papers  published in 1966, V. I. Arnold changed the way we think about fluid dynamics, forever. In the papers \cite{Arnold:1966aa, Arnold:1966ab}, Arnold showed that the solutions of the Euler fluid equations in a domain $M$ in fixed space, $M\subset \mathbb{R}^n$, can be mapped by the classic Lagrange-Euler representation to a time-dependent path on the manifold of volume-preserving diffeomorphisms acting on $\mathbb{R}^n$ (SDiff$(\mathbb{R}^n)$) which is geodesic with respect to the right-invariant metric on its tangent space given by the kinetic energy of the fluid. The kinetic energy metric is right-invariant because it is the $L^2$ norm of  the right-invariant Eulerian velocity $u_t$ defined by
\begin{align}\label{Eul-vel-def}
\dot{g}_t x_0 = u_t({g}_t x_0)
\,,
\end{align}
in which subscript $t$ denotes explicit  time dependence.  This $g_t$ is  simply the Lagrangian flow map at time $t$ associated to the velocity field $u$.
The Lagrangian fluid parcel trajectory is given by
\begin{align}\label{Lag-traj-def}
x_t:={g}_t x_0\in M
\quad\hbox{with}\quad
{g}_0 x_0=x_0
\,.
\end{align}
Upon writing $u_t = \dot{g}_t{g}_t^{-1}$ one sees that the invariance of the Eulerian velocity  corresponds to relabelling the Lagrangian parcel label  $x_0\to y_0=hx_0$ for a fixed map $h\in {\rm SDiff}(\mathbb{R}^n)$. Thus, under any fixed map $h$ acting from the right we have $\dot{g}_th(g_th)^{-1}=\dot{g}_t{g}_t^{-1}$. Arnold's identification of the Euler fluid solutions as geodesics also brings in Hamilton's variational principle, in which right-invariance summons Noether's theorem for Lie group invariant variational principles. 

Arnold's idea that Euler fluid flows could be lifted to time-dependent paths on SDiff$(\mathbb{R}^n)$ has been continually fruitful. Already in 1970, Ebin and Marsden \cite{Ebin:1970aa} used this idea to prove the local in time existence and uniqueness of the Euler fluid flows in $\mathbb{R}^3$. This is still the definitive analytical result for the Euler fluid equations. By 1985, Marsden and his collaborators had used the same idea to obtain the Lie--Poisson Hamiltonian formulation for ideal fluids with advected quantities and had recognised the role of its semidirect-product structure in establishing nonlinear stability for a wide class of  fluid and plasma equilibria for continuum flows with advected quantities and additional physical fields \cite{Holm:1983aa, Holm:1985aa}. Again, this development of nonlinear stability conditions followed Arnold's lead in \cite{Arnold:1966aa} for the nonlinear stability of Euler fluid equilibria as critical points of a constrained variational principle for time dependent paths on the manifold SDiff$(\mathbb{R}^n)$.  For further explanation of these developments in the context of momentum maps, see \cite{Marsden:2001aa,Weinstein:2005aa}.

Following an observation reported in 1901 by Poincar\'e,  \cite{Poincare:1901aa}, Holm, Marsden and Ratiu \cite{Holm:1998aa} transferred the idea of symplectic reduction for Hamiltonian systems into the theory of reduction by Lie symmetries of the Lagrangian in Hamilton's principle and applied the resulting Euler--Poincar\'e equations to derive the general dynamics of fluid flows with advected quantities. This result again followed Arnold's lead in regarding fluid flows as curves on SDiff$(\mathbb{R}^3)$, although the dynamics discussed in \cite{Holm:1998aa} takes place on Diff$(\mathbb{R}^3)$ in the compressible case when volume is not preserved. An interesting feature in this particular development involves the Kelvin--Noether theorem. This theorem revealed that the momentum map on the Lagrangian side for the action corresponding to the relabelling symmetry was, in fact, the circulation integral in Kelvin's theorem for Euler's equations. Thus, the conservation of the Kelvin circulation integral for the Euler fluid equations was found to arise via Noether theorem from the symmetry of the Eulerian velocity in \eqref{Eul-vel-def} under relabelling of the Lagrangian fluid parcels. 

Now, the Lagrangian fluid parcels carry advected physical variables such as mass, heat, other thermodynamic quantities, buoyancy for stratification of fluid motion under gravity, magnetic field lines for MHD, etc.  The introduction of spatially varying initial conditions for these advected quantities breaks the symmetry of the Euler kinetic energy Lagrangian under the full set of Lie group transformations by $G={\rm Diff}(\mathbb{R}^n)$. In particular, the invariance of the Lagrangian is restricted to the ``isotropy subgroups''  $G_{a_0}:={\rm Diff}_{a_0}(\mathbb{R}^n)$ of the full Diff$(\mathbb{R}^n)$ Lie transformations. The isotropy subgroups $G_{a_0}$ are those which leave invariant the initial conditions $a_0$ chosen for the advected quantities. The advected quantities evolve by push-forward $a_t = a_0g_t^{-1}$ by the action of the entire Diff$(\mathbb{R}^n)$. (Push-forward is pull-back by right action of $g_t^{-1}$.) Thus, symmetry breaking from Diff$(\mathbb{R}^n)$ to Diff$_{a_0}(\mathbb{R}^n)$ leads to the identification of advected quantities as order parameters $a_t\in G/G_{a_0}$, where $G/G_{a_0}$ is the corresponding coset space of the broken symmetry under the action of $G$ by the remaining symmetry $G_{a_0}$ for $a_0$. This symmetry breaking of Diff$(\mathbb{R}^n)$ for the Euler fluid equations implies that the Euler--Poincar\'e equations for flows with advected quantities acquire force terms of geometric origin, thereby leading to the so-called ``diamond $(\,\diamond\,)$ terms'' which will be discussed further below. Including these force terms arising from symmetry breaking implies that fluid solutions with advected quantities are no longer geodesic paths on Diff or SDiff. For an application of these ideas to complex fluids such as liquid crystals, see \cite{Holm:2002aa, Gay-Balmaz:2010aa}.

This history of the development of Arnold's SDiff flow concept and its extension to include advected fluid quantities provides the context of flows on Diff or SDiff  for the introduction of the material addressed in the present paper. Because of its close connection to Lie symmetry transformations via Noether's theorem, this material can be addressed operationally and quite transparently from the viewpoint of Kelvin's theorem, by using pull-back by the time dependent flow. This operational interpretation arises naturally because of the physical connection of Kelvin's theorem to Newton's force law. Namely, the Kelvin circulation integral is the Noether quantity describing momentum distributed on closed material loops \cite{Holm:1998aa}.

\section{Geometric setting}\label{sec:GeomSet}

In this section, we define the elements of the geometric/probabilistic framework which facilitate the interpretation of  our equations.

\vspace{1mm}\noindent \textbf{Basic setting and notation.}
Let $(M,g)$ be a $C^{\infty}$, compact, oriented, $d$-dimensional Riemannian manifold with empty boundary $\partial M= \emptyset$. We will denote by $dV$ the Riemannian volume form on $M$ associated with the metric. 

Let $\mathfrak{X}$ denote the space of smooth vector fields on $M$,  $\mathcal{T}^{r}_s$ denote the space of smooth $r$-contravariant $s$-covariant tensor fields  on $M$ with  the tensor product denoted by $\otimes$. Let $\Omega^k$ denote the space of smooth exterior $k$-forms on $M$ with  the wedge product denoted $\wedge$. Let  $\mathbf{d}:\Omega^k \rightarrow \Omega^{k+1}$ denote the exterior differential operator and $\mathbf{i}_{u}: \Omega^{k+1}\rightarrow \Omega^k$ the interior product  for a given $u\in \mathfrak{X}$, both of which are antiderivations.

For a given vector field $u\in \mathfrak{X}$ and $\tau\in\mathcal{T}^{r}_s$ , let $\pounds_{X}\tau$ denote the Lie derivative of $\tau$ along $X$, which can be defined by $\pounds_{X}\tau=\frac{d}{dt}|_{t=0}\phi^*_{t}\tau$, where for each $t$, $\phi^*_t\tau$ denotes the pull-back of the tensor field $\tau$ by the flow map $\phi_t$ associated with $u$. 

For a given tensor field $\alpha\in \Omega^k$ and $u\in \mathfrak{X}$, the Lie derivative satisfies Cartan's formula
\[\pounds_{u} \alpha=\mathbf{d}(\mathbf{i}_u \alpha) + \mathbf{i}_u\mathbf{d}\alpha\,.\]  
Moreover,  in a local  chart $(U,\phi)$ of $M$ with coordinates $(x^1,\ldots, x^d)$, one has
$$
\pounds_u v =-\operatorname{ad}_uv= [u,v]=(u^j\partial_{x^j}v^i - v^j\partial_{x^j}u^i)\partial_{x^i}
, \quad\hbox{for}\quad 
u=u^i\partial{x^i},\ v=v^i\partial_{x^i}\in \mathfrak{X}|_{U}\,,
$$
as well as
$$
(\pounds_{u}\omega)_{i_1\cdots i_k}= u^j\partial_{x^j}\omega_{i_1\cdots i_k} + \omega_{j\cdots i_k}\partial_{x^{i_1}}u^j +\cdots +\omega_{i_1\cdots j}\partial_{x^{i_k}}u^j,
$$
$$
\hbox{for}\quad \omega=\omega_{i_1\cdots i_k}dx^{i_1}\wedge \cdots \wedge dx^{i_k}\in \Omega^k|_{U}, \;i_1<\cdots <i_k,
$$
and
$$
(\pounds_{u}\tau)^{j_1\cdots j_r}_{i_1\cdots i_s}= u^j\partial_{x^j}\tau_{i_1\cdots i_s}^{j_1\cdots j_r} + \tau_{j\cdots i_s}^{j_1\cdots j_s}\partial_{x^{i_1}}u^j +\cdots +\tau_{i_1\cdots j}^{j_1\cdots j_r}\partial_{x^{i_s}}u^j+\tau_{i_1\cdots i_s}^{j_1\cdots j_r}\partial_{x^{j_1}}u^j +\cdots +\tau_{i_1\cdots i_k}^{j_1\cdots j}\partial_{x^{j_r}}u^j,
$$
$$
\hbox{for}\quad \tau = \tau^{j_1\cdots j_r}_{i_1\cdots i_s}\partial_{x^1}\otimes \cdots \otimes \partial_{x^r}\otimes dx^{i_1}\otimes \cdots\otimes dx^{i_s}\in \mathcal{T}^r_s|_{U}.
$$

\vspace{1mm}\noindent \textbf{Musical isomorphisms, the Hodge star operator, and the Hodge decomposition.} The metric $g$ induces an isomorphism between $\mathfrak{X}$ and $\Omega^1$; we denote by $\flat:\mathfrak{X} \rightarrow \Omega^1$ the index lowering map and by  $\sharp: \Omega^1\rightarrow \mathfrak{X}$  the index raising map. Similarly,  the indices of a  tensor field or differential form can be raised or lowered by using the metric. Furthermore, the metric $g$ automatically induces Riemannian structures on the bundles $\mathcal{T}^r_s$  and $\Omega^k$, which we continue to denote by $g$. Indeed, for $\alpha,\beta \in \Omega^1$,  $g(\alpha,\beta)=\alpha(\beta^{\sharp})=\beta(\alpha^{\sharp})=g(\alpha^{\sharp},\beta^{\sharp}),$ and then the metric on $\mathcal{T}^r_s$ is defined by the tensor product $\otimes^rg \otimes \otimes^s g$.

The metric  and  orientation uniquely determines the Hodge star operator $\star : \Omega^{k}\rightarrow \Omega^{d-k}$, which satisfies  $\alpha \wedge \star \beta = g(\alpha, \beta)\mu$ for all $\alpha,\beta\in \Omega^k$. The corresponding co-differential $\delta: \Omega^k\rightarrow \Omega^{k-1}$ is defined by $\delta = (-1)^{d(k+1)+1}\star \mathbf{d}\star$. The metric also induces a pre-inner product on $\mathcal{T}^r_s$ (and hence on $\Omega^k$) defined by
\begin{equation}\label{eq:aninner}
(\tau,\tau')_{L^2(\mathcal{T}^r_s)}=\int_{M} g(\tau,\tau')dV, 
\end{equation}
for $ \tau,\tau'\in \mathcal{T}^r_s.$ It can be shown that for all $\alpha,\beta \in \Omega^k$, 
$$
(\alpha,\beta)_{L^2(\Omega^k)} = \int_M g(\alpha,\beta)dV=\int_M g(\alpha,\beta)\star 1=\int_M \alpha \wedge \star \beta\,,
$$
and that the co-differential $\delta$ is formally the adjoint of the differential $\mathbf{d}$ on $\cup_{k=0}^d \Omega^k$ with respect to this inner product.

Letting $L^2(\Omega^k)$ denote the space of Borel measurable square-integrable equivalent classes of functions $\Omega^k$ and  equipping this space the  inner product $(\cdot,\cdot)_{L^2(\Omega^k)}$ produces a Hilbert space.  The Hodge decomposition (see e.g., Ch. 3 of \cite{Jost:2008aa})  associated with the Hodge Laplacian $\Delta = \mathbf{d}\delta +\delta \mathbf{d}$, then reads 
$$
L^2(\Omega^k)=\overline{\delta\Omega^{k+1}}\oplus\overline{\mathbf{d}\Omega^{k-1}}\oplus  \mathcal{H}^k ,
$$
where $\overline{(\,\cdot\,)}$ denotes the closure in $L^2(\Omega^k)$ of the corresponding set and $\mathcal{H}_k=\{\alpha\in \Omega^k : \Delta \alpha=0\}$ is a space of harmonic-$k$ forms, which is  orthogonal to the other two components.   Noting the identity $\mathbf{i}_{u}\mu = \star u^{\flat}$ for $u\in \mathfrak{X}$, we find that $\operatorname{div}_g(u)=-\delta u^{\flat}$, where  $\operatorname{div}_g(u)$ is defined to be the unique element of $\Omega^0=C^{\infty}(M)$ such that 
$$
\pounds_{u} dV = \operatorname{div}_g(u)dV=\mathbf{d}(i_{u}dV)\,. 
$$ 
Therefore, the Hodge decomposition of $L^2(\Omega^1)$ says that for every $u\in \mathfrak{X}$,  
$$
u^{\flat} = \mathbf{P}u^{\flat}+\mathbf{Q}u^{\flat}+\mathbf{P}_{\mathcal{H}}u^{\flat}, 
$$
where the sharp of $\mathbf{P}u^{\flat}$ is a weakly divergence-free vector field, the sharp of  $\mathbf{Q}u^{\flat}=\mathbf{d} p$ is the weak-gradient of a scalar function  and the sharp of $\mathbf{P}_{\mathcal{H}}u^{\flat}$ is the harmonic-form part. It can also be shown that spaces $W^{m,p}(\Omega^k)$ or $C^{\alpha}(\Omega^k)$ of Sobolev or H\"older differential forms, respectively, admit Hodge decompositions in which  there respective divergence-free or gradient-components have the same corresponding regularity (see, e.g., \cite{morrey2009multiple,palais1968foundations}).

\vspace{1mm}\noindent\textbf{Momentum, one-form densities, weak duals, ad-star and ad-dagger.}
The LA SALT theories presented in the text are based on Euler-Poincare (EPDiff) theory \cite{Holm:1998aa}, in which the (weak) dual spaces  of $\mathfrak{X}$, $\Omega^k$,  $\mathcal{T}^r_s$, and $\mathcal{T}^r_s \otimes \Omega^d$ with respect to  $L^2$-pairings play a prominent role. This is because the theory is founded on variational principles in which Gateux derivatives of functionals (e.g., the Lagrangian and constraints) involve $L^2$-norms (see \eqref{def:GateuxLag}).

Since we have fixed a metric $g$,  a natural candidate for the weak duals of $\mathfrak{X}$, $\Omega^k$,  $\mathcal{T}^r_s$, and $\mathcal{T}^r_s \otimes \Omega^d$, respectively, can be defined using the $L^2$-pairing \eqref{eq:aninner} to be $\mathfrak{X}$, $\Omega^k$,  $\mathcal{T}^r_s$, and  $\mathcal{T}^r_s \otimes \Omega^d$, respectively. That is, these spaces are weakly dual to themselves using the $L^2$-inner-product (based on the metric). However, the geometric structure and conservation laws such as Kelvin's theorem and its generalization are  more readily apparent if one defines weak ``geometric duals'' as defined below and writes the Euler-Poincar\'e equations in terms of these geometric dual spaces. It is important to emphasize that the theory does not depend on how the duals are defined since one can always transform the equations among the various definitions.

Let $\operatorname{Den}= \{ \nu \in \Omega^d:  \nu=\rho dV,\;\rho \in \Omega^0\} \subset \mathcal{T}^0_{d+1}$. We identify the weak (geometric) dual of $\mathfrak{X}$ with  $\mathfrak{X}^*=\Omega^1\otimes \operatorname{Den}$ via the pairing $\langle \cdot, \cdot\rangle_{\mathfrak{X}}: \mathfrak{X}^*\times \mathfrak{X}\rightarrow \mathbb{R}$ defined by
\begin{equation}\label{def:weak_dual_vec}
\scp{\mu}{u}_{\mathfrak{X}}:=\int_{M}\alpha(u) \rho dV,\;\;  \mu=\alpha \otimes \rho dV \in \mathfrak{X}^*, \;\;   u\in \mathfrak{X}.
\end{equation}
It follows that the formal adjoint of $\operatorname{ad}_{v}:\mathfrak{X}\rightarrow \mathfrak{X}$ relative to the pairing $\langle \cdot, \cdot\rangle_{\mathfrak{X}}$  is $\operatorname{ad}_{v}^*=\pounds_{v}$; that is, for all $\mu \in \mathfrak{X}^*$ and $u,v\in \mathfrak{X}$, 
$$
\langle \mu  , \operatorname{ad}_vu \rangle_{\mathfrak{X}} = \langle \pounds_v \mu ,  u\rangle_{\mathfrak{X}}.
$$
Indeed, owing to the definition of the Lie-derivative, the change of variable formula, and the natural properties of the tensor product and contraction, for all $\alpha \otimes \rho dV\in\mathfrak{X}^*$ and $u,v\in \mathfrak{X}$, we have
\begin{align*}
\langle \alpha \otimes \rho dV, \operatorname{ad}_vu\rangle_{\mathfrak{X}}&=\int_{M}\alpha(\operatorname{ad}_v u) \rho dV=-\frac{d}{d\epsilon}\big |_{\epsilon=0} \int_{M}\alpha(\phi^*_{\epsilon} u) \rho dV\\
&=-\frac{d}{d\epsilon}\big |_{\epsilon=0} \int_{M}((\phi_{\epsilon})_*\alpha)(u) (\phi_{\epsilon})_*(\rho dV)\\
&=-\int_{M}\left[(\pounds_{-v}\alpha)(u)\rho dV + \alpha(u) \pounds_{-v}( \rho dV)\right]\\
&=\langle \pounds_{v}\left(\alpha \otimes \rho dV\right), u\rangle_{\mathfrak{X}},
\end{align*}
in which $\{\phi_{\epsilon}\}$ is the flow generated by $v$. 
Moreover, for all $\mu \in\mathfrak{X}^*$ and $u,v\in \mathfrak{X}$, we have
\begin{equation}\label{eq:LieAdRelations}
\langle \pounds_{u}\mu , v\rangle_{\mathfrak{X}}=\langle \mu  , \operatorname{ad}_uv\rangle_{\mathfrak{X}}=-\langle \mu , \operatorname{ad}_vu\rangle_{\mathfrak{X}}=-\langle \pounds_{v}\mu , u\rangle_{\mathfrak{X}}. 
\end{equation}
In geometric fluid dynamics, the momentum $\mu=\delta\ell/\delta u$ belongs to $\mathfrak{X}^*$. 

\vspace{1mm}\noindent\textbf{Adjoint operations.}
The formal adjoint of $\operatorname{ad}_{v}:\mathfrak{X}\rightarrow \mathfrak{X}$ with respect to the $L^2(\mathfrak{X})$-inner-product is denoted by $\operatorname{ad}_v^{\dagger}$. That is, 
$$ (\operatorname{ad}_vu,w)_{L^2(\mathfrak{X})}=(u,\operatorname{ad}_v^{\dagger}w)_{L^2(\mathfrak{X})}, \;\; \forall u,v,w\in \mathfrak{X}.
$$
It follows that for all $u,v,w\in \mathfrak{X}$,
$$
(\operatorname{ad}_v^{\dagger}u,w)_{L^2(\mathfrak{X})}=\Scp{ \pounds_{v}\left(u^{\flat}\otimes dV\right)}{w}_{\mathfrak{X}}=\Scp{ \operatorname{ad}^*_v\left(u^{\flat}\otimes dV\right)}{w}_{\mathfrak{X}},
$$
and hence for all $\alpha\in \Omega^1$,
\begin{equation}\label{eq:addagger}
\operatorname{ad}_v^{\dagger} u = (\pounds_{v} u^{\flat})^{\sharp}+(\operatorname{div}_gv) u, \quad 
\operatorname{ad}_v^{\dagger} \alpha^{\sharp} = (\pounds_{v} \alpha)^{\sharp}+(\operatorname{div}_gv) \alpha^{\sharp}.
\end{equation}\
Moreover, for all $u,v,w\in \mathfrak{X}$, we have
\begin{equation}\label{eq:addaggerrel}
(\operatorname{ad}^{\dagger}_v u, w)_{L^2(\mathfrak{X})}=(u, \operatorname{ad}_v w)_{L^2(\mathfrak{X})}=-(u, \operatorname{ad}_w v)_{L^2(\mathfrak{X})} =-(\operatorname{ad}^{\dagger}_w u, v)_{L^2(\mathfrak{X})}.
\end{equation}

For a given $\mu \in \mathfrak{X}^*$ and $D = \rho dV\in \operatorname{Den}$, we define the one-form $\frac{\mu}{D} \in \Omega^1$  by expressing the one-form density $\mu$ in terms of the volume-form $ dV$ as
\[
\mu = \frac{\mu}{D} \otimes D = \frac{\mu}{D} \otimes \rho dV\,.
\]

\vspace{1mm}\noindent\textbf{Advected variables and the diamond operation.} Let us define the weak (geometric) dual of  $\Omega^{k}$ to be  $\Omega^{d-k}$ via  the pairing 
$$
\langle \beta, \alpha\rangle_{\Omega^{k}}=\int_{M} \alpha \wedge \beta ,\;\; \alpha\in \Omega^k, \  \beta \in \Omega^{d-k}.
$$ 
Let $V=\operatorname{Den}\oplus \oplus_{i=1}^N\Omega^{k_i}$ for $k_i\in \{0,\ldots, d\}$, which is the space in which the advected quantities will live.  We denote by $V^*$  the (geometric) weak dual of $V$ relative to the  weak non-degenerate pairing $\langle \cdot, \cdot\rangle_V: V^*\times V\rightarrow \mathbb{R}$ defined by
$$\langle \cdot, \cdot \rangle_{V}=\langle \cdot, \cdot\rangle_{\Omega^d}+\sum_{i=1}^N\langle \cdot, \cdot\rangle_{\Omega^{k_i}}.$$ It follows that for all $a=\{a^{(i)}\}_{i=0}^N, b=\{b^{(i)}\}_{i=0}^N\in V$,  
\begin{equation}\label{eq:advectedsym}
(a,b)_{L^2(V)}:=\sum_{i=0}^N(a^{(i)}, b^{(i)})_{L^2(\Omega^{k_i})}^2=\sum_{i=0}^N\int_{M}a^{(i)}\wedge \star b^{(i)}=\langle \star b, a\rangle_{V}=\langle \star a, b\rangle_{V}.
\end{equation}
One can define geometric duals for tensor or tensor densities using tensor contraction and allow  advected variables in $V$ to lie in these spaces as well. However, for simplicity, we will only treat the case of $k$-forms. 

Let $\diamond : V^*\times V\rightarrow \mathfrak{X}^*$ be the  bi-linear operation defined by the relation
\begin{align}\label{diamond-def}
\scp{b\diamond a}{u}_{\mathfrak{X}} = \scp{b}{- \pounds_u a}_V=\sum_{i=0}^N\scp{b_i}{-\pounds_{u}a^{(i)}} \,, \;\; \forall a=\{a^{(i)}\}_{i=0}^N\in V, \; b\in V^*, \; \forall u\in \mathfrak{X}.
\end{align}
The diamond can be calculated on a case-by-case basis using the definition of the Lie-derivative (and Cartan's formula), Stoke's theorem, and the antiderivation properties of the exterior differential and insertion operators.  See \S \ref{sec:LASALT-Eu} and  \S \ref{sec:examples} for concrete examples including LA SALT Euler and MHD.  

We define $\hat{\diamond} : V^*\times V \rightarrow \mathfrak{X}$ by $b\, \hat{\diamond} \, a = \left(\frac{1}{dV}\left(b\diamond a\right)\right)^{\sharp}$ for all $a\in V$, $b\in V^*$. It follows that for all $a\in V, b\in V^*$ and $u \in \mathfrak{X}$, 
\begin{equation}\label{eq:hat_diamond}
(b\,\hat{\diamond}\,  a,\, u)_{L^2(\mathfrak{X})}=\int_{M} \left(\frac{1}{dV}(b\,\hat{\diamond}\,  a)\right)(u)dV=\langle b \diamond a, u\rangle_{\mathfrak{X}}=\langle b, -\pounds_{u} a\rangle_{V}.
\end{equation}
\end{appendices}

\subsection*{Acknowledgments.} We are grateful for stimulating and encouraging discussions of the topics in this paper with C. J. Cotter, D. Crisan, A. Bethencourt de Leon, T.\ Nilssen and S. Takao. We particularly thank S. Takao for helpful discussions on the fluctuation variance dynamics in \S \ref{sec:FluctDyn}. Research of TD is partially supported by NSF-DMS grant 1703997. DDH and JML are grateful for partial support by the EPSRC Standard Grant EP/N023781/1.

\newpage 
\bibliographystyle{unsrt}
\bibliography{bibliography}
\end{document}